\documentclass[showpacs,prl,onecolumn,aps,superscriptaddress,preprintnumbers,nofootinbib]{revtex4}
\usepackage[T1]{fontenc}
\usepackage[latin9]{inputenc}
\usepackage{amsmath,amssymb}
\usepackage{epsfig}
\usepackage{graphicx}
\usepackage{mathrsfs}
\usepackage{amsmath}
\usepackage{amsfonts}
\usepackage{epstopdf}
\usepackage{color}
\def\slashchar#1{\setbox0=\hbox{$#1$}     		
   \dimen0=\wd0                                 	
   \setbox1=\hbox{/} \dimen1=\wd1               	
   \ifdim\dimen0>\dimen1                        	
      \rlap{\hbox to \dimen0{\hfil/\hfil}}      	
      #1                                        	
   \else                                        	
      \rlap{\hbox to \dimen1{\hfil$#1$\hfil}}   	
      /                                         	
   \fi}

\renewcommand{\vec}{\boldsymbol}
\newcommand{\be}{\begin{equation}}
\newcommand{\ee}{\end{equation}}
\newcommand{\bea}{\begin{eqnarray}}
\newcommand{\eea}{\end{eqnarray}}
\newcommand{\ba}{\begin{array}}
\newcommand{\ea}{\end{array}}

\def\eq#1{{Eq.~(\ref{#1})}}
\def\fig#1{{Fig.~\ref{#1}}}
\newcommand{\bas}{\bar{\alpha}_S}

\newcommand{\nn}{\nonumber}

\newcommand{\Lb}{\left(}
\newcommand{\Rb}{\right)}
\newcommand{\h}{\frac{1}{2}}

\begin{document}

\title{ High energy QCD: multiplicity distribution and entanglement entropy}
\author{E. ~Gotsman}
\email{gotsman@post.tau.ac.il}
\affiliation{Department of Particle Physics, School of Physics and Astronomy,
Raymond and Beverly Sackler
 Faculty of Exact Science, Tel Aviv University, Tel Aviv, 69978, Israel}
 \author{E.~ Levin}
\email{leving@tauex.tau.ac.il, eugeny.levin@usm.cl}
\affiliation{Department of Particle Physics, School of Physics and Astronomy,
Raymond and Beverly Sackler
 Faculty of Exact Science, Tel Aviv University, Tel Aviv, 69978, Israel}\affiliation{ Departamento de F\'\i sica,
Universidad T$\acute{e}$cnica Federico Santa Mar\'\i a   and
Centro Cient\'\i fico-Tecnol$\acute{o}$gico de Valpara\'\i so,
Casilla 110-V,  Valparaiso, Chile}

\date{\today}


\begin{abstract}
In this paper we show that QCD at high energies leads to the
 multiplicity distribution $\frac{\sigma_n}{\sigma_{ \rm in}}\,\,=
\,\,\frac{1}{N}\,\Lb \frac{N\,-\,1}{N}\Rb^{n - 1}$,   (where $N$ denotes
 the average number of 
particles), and to 
entanglement entropy $S \,=\,\ln N$, confirming that the partonic state
 at high energy is maximally entangled. However, the value of $N$ depends
 on the kinematics of the parton cascade. In particular, for DIS 
 $N = xG(x,Q)$ , where $xG$ is the gluon structure function, while
 for hadron-hadron collisions, $N \propto Q^2_S(Y)$, where $Q_s$ denotes
 the saturation scale. We checked that this multiplicity distribution
 describes the LHC data for low multiplicities $n \,<\,(3 \div 5)\,N$,
  exceeding it  for larger values of $n$. We view this as a  result
 of our
 assumption, that the  system of partons in hadron-hadron collisions at
 c.m. rapidity $Y=0$ is dilute. We show that the data can be
 described at large multiplicities in the parton model,  if we do not 
make   this assumption.
\end{abstract}
\maketitle

\vspace{-0.5cm}
\tableofcontents

\section{Introduction}
 Over  the past several years  new ideas have been developed in 
the high
 energy and nuclear physics community, which suggest a  robust 
relation
 between the principle features of high energy scattering and entanglement
 properties of the hadronic wave function
\cite{KUT,PES,KOLU1,PESE,KHLE,BAKH,BFV,HHXY,KOV1,GOLE1,GOLE2,KOV2,
NEWA,LIZA,FPV,TKU,KOV3}. The main idea, which we explore in this
 paper,  is the intimate relation between the entropy in the
 parton approach\cite{BJ,FEYN,BJP,Gribov} and  the entropy of
 entanglement in a proton wave function\cite{KHLE}. 

This relation  materialized   as the resolution of the following 
difficulty in 
 our understanding of high energy scattering: on one hand, the proton
 is  a pure
state and it is described by a completely coherent wave function with
 zero entropy, but, on the other hand, the DIS experiments are successfully
 described, treating the proton as a  incoherent collection of quasi-free
 partons. This ensemble  has non vanishing entropy, and Ref.\cite{KHLE}
 proposes, that the origin of this entropy is the 
 entanglement between the degrees of freedom one observes in DIS (partons
 in the small spatial region of
the proton), and the rest of the proton wave function, which is not 
measured
 in the DIS experiments.

 In other words,  the hadron in the rest frame is described by a  pure 
state 
 $|\psi\rangle$ with density matrix $\hat{\rho} = |\psi\rangle \langle
 \psi |$ and zero von Neumann entropy $S = - \rm{tr} \left[ \hat{\rho}
 \ln \hat{\rho} \right] = 0$. In DIS 
 at Bjorken $x$ and momentum transfer $q^2 = - Q^2$ probes only a part
 of the proton's wave function; let us denote it $A$.
   In the proton's rest frame the DIS probes the spatial region $A$
 localized within a tube of radius $\sim 1/Q$ and length $\sim 1/(mx)$
 \cite{GIP,Ioffe}, where $m$ is the proton's mass. The inclusive DIS
 measurement thus sums over the unobserved part of the wave function
 localized in the region $B$ complementary to $A$, so we have access
 only to the reduced density matrix $\hat{\rho}_A = \rm{tr}_B \hat{\rho}$,
 and not the entire density matrix $\hat{\rho} = |\psi\rangle \langle
 \psi |$.  In Ref.\cite{KHLE} it is proposed that
     \be \label{I1}
   S_A = - \rm{tr}_B \left[ \hat{\rho}_A \ln \hat{\rho}_A \right]\,\,=\,\, S_{\rm parton \,cascade}
   \ee
\eq{I1}, in spite of its general form,  means that we can 
estimate  the
 entropy and multiplicity distribution of the produced gluons using the
 parton wave function in the initial state.  In addition , we can 
obtain a
 thermal distribution of the produced particles in the high energy
 collision in spite of the fact, that
 the number of  secondary interactions in proton-proton collisions
 is rather low, and cannot provide the thermalization due to the
 interaction in the final state.  
 
 It has been demonstrated  in Refs.\cite{BAKH,GOLE1,GOLE2,TKU}  
 that these 
ideas are in qualitative and, partly, in  quantitative  agreement with the
 available experimental data.

The goal of this paper is to study the multiplicity distribution and the
 entanglement entropy in the effective theory for  QCD at high energies
 (see Ref.\cite{KOLEB} for a general review). Such a theory exists  
in two different formulations:  the CGC/saturation approach
 \cite{MV,MUCD,B,K,JIMWLK,GIJMV}, and the BFKL Pomeron calculus 
\cite{BFKL,LI,GLR,GLR1,MUQI,MUPA,BART,BRN,KOLE,LELU1,LELU2,LMP,
AKLL,AKLL1,LEPP}. 

We believe that 
the CGC/saturation approach  provides  a more 
general
  pattern \cite{AKLL,AKLL1}   for the treatment of high energy 
QCD. However, in this paper we restrict ourself   to  the BFKL 
Pomeron
 calculus, which has a more direct correspondence with the parton 
approach,  and has  been used in Ref.\cite{KHLE}.
  
  Fortunately, in Ref.\cite{AKLL1} it was shown, that 
these
 two approaches are equivalent for the description of the scattering amplitude
\be \label{I2}
Y \,\leq\,\frac{2}{\Delta_{\mbox{\tiny BFKL}}}\,\ln\Lb
 \frac{1}{\Delta^2_{\mbox
{\tiny BFKL}}}\Rb
\ee
where $\Delta_{\mbox{\tiny BFKL}}$ denotes the intercept of the BFKL 
 Pomeron.

The main difference between the CGC approach and the parton QCD 
 cascade for the  topics dealt with  in this paper, is the fact that the
 CGC approach generates the non-diagonal elements of the density matrix
 (see for example Refs.\cite{KOLU1,KOV1,KOV2,KOV3}), while for  the
 parton cascade and, generally, in the BFKL Pomeron calculus the
 density matrix is diagonal.  Since in DIS experiments we can only
 measure  the diagonal elements of density matrix, we can
 introduce in the framework another kind of entropy:`` the entropy
 of ignorance"\cite{KOV3},   which characterized this lack of knowledge
 of the actual density matrix  in DIS experiments.  We will show below
 that the McLerran-Venugopalan approach\cite{MV} ,that is used in 
Ref.\cite{KOV3}, leads to the same multiplicity distribution as the
 parton cascade.

The paper is organized as follows: In the next section we
 consider the entropy and multiplicity distributions  in the QCD parton
 cascade. We show that in spite of the fact that in different kinematic
 regions,  the QCD cascade leads to a different energy and dipole size
 dependence of the mean multiplicity, and  the multiplicity distribution 
has
 a general form:
\be \label{I3}
\frac{\sigma_n}{\sigma_{ \rm in}}\,\,=\,\,\frac{1}{N}\,\Lb \frac{N\,-\,1}{N}\Rb^{n - 1}\,
\ee
 where N is the
 average number of partons. The entanglement entropy is
 equal to $S_{\rm parton \,cascade}\,\,=\,\,\ln N$,  confirming that
 {\it  the partonic state
 at high energy is maximally entangled} \cite{KHLE}. In the case of
 DIS we argue that $N$ is equal to the  gluon structure function,  but 
  we can only prove the 
 multiplicity distribution of \eq{I3},   for the BFKL
 evolution of this structure function. In section III we show that the CGC
 approach leads  to the multiplicity distribution of \eq{I3}. In section 
IV
 we consider  hadron-hadron scattering. In the range of energy given
 by \eq{I2} we use the Mueller-Patel-Salam-Iancu(MPSI) \cite{MUPA,MPSI}
 approach, using the formalism of Ref.\cite{LELU2}. We show that in the
 framework of this approach we have the distribution of \eq{I3} 
which can describe the experimental data for sufficiently 
 low multiplicities $n \leq (3\div 5)<n>$ . However, we fail
 to describe   the data for  larger $n$. We conclude that the main 
assumption
 of the MPSI approach, that a system of dilute  partons are produced
 in the c.m. rapidity Y=0, is not valid  for large multiplicities
  at high energies of the LHC. Unfortunately, at the moment we 
have no theoretical tool to treat this scattering. However,
 in Ref.\cite{KLL} an  approach has been suggested, which allows
 us to describe the dense system of partons in hadron-hadron collisions,
 as well as the dilute one. Developing this approach for the multiplicity 
distribution,  we  are able to describe the data for large 
multiplicities. 
 We summarize our results in the conclusions.

\section{The QCD parton cascade}
\subsection{QCD cascade for fast moving large dipole}

\subsubsection{General approach}

\begin{figure}[ht]
     \begin{center}
     \includegraphics[width=1\textwidth]{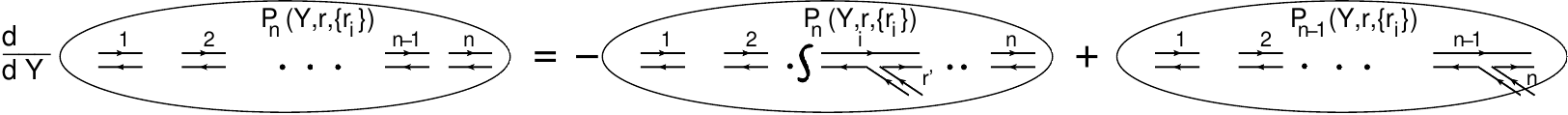} 
     \end{center}    
      \caption{ The graphical form of \eq{PC1}.}
\label{eq}
   \end{figure}

 As discussed in Refs.\cite{KOLEB,MUCD,LELU1,LELU2}  the parton cascade can
 be written  in the following form (see \fig{eq}):
  \bea  \label{PC1}
&&\frac{\partial\,P_n\left(Y, \vec{r }, \vec{b};\,\vec{r}_1,\vec{ b}_1,\,\vec{r}_2 , \vec{b}_2\dots \vec{r}_i ,\vec{b}_i,
\dots \vec{r}_n, \vec{b}_n \right)}{ 
\partial\, Y }\,=\,-\,
\sum^n_{i=1}\,\omega_G(r_i) \,
P_n\left(Y, \vec{r }, \vec{b};\,\vec{r}_1,\vec{ b}_1,\,\vec{r}_2 , \vec{b}_2\dots \vec{r}_i ,\vec{b}_i,
\dots \vec{r}_n, \vec{b}_n \right) \\
&&~~~~~~~~~~~~~~~~~~~~~~~~~~~~~~~~~~~~~~~~~~~~~+\,\,\bas\,\sum^{n-1}_{i=1} \,\frac{(\vec{r}_i\,+\, 
\vec{r}_n)^2}{(2\,\pi)\,r^2_i\,r^2_n}\,
P_{n - 1}\left(Y, \vec{r},\vec{b};\,\vec{r}_1,\vec{b}_1,
\dots  (\vec{r}_i \,+\, \vec{r}_n), \vec{b}_{in},\dots \vec{r}_{n-1},\vec{b}_n
 \right)\nn
\eea
  where $P_n\Lb Y ; \{r_i,b_i\}\Rb$ is the probability to have $n$-dipoles
 of size $r_i$,  at impact parameter $b_i$ and  at rapidity $Y$\footnote
{ In the lab. frame rapidity $Y$ is equal to   $Y = y_{\rm dipole ~r} \,-\,y_{\rm dipoles~r_i}$, where $y_{\rm dipole ~r}$ is the rapidity of the incoming fast dipole and $y_{\rm dipole~r_i}$ is the rapidity of dipoles $r_i$.}
  . $\vec{b}_{in} $ in \eq{PC1} is equal to $\vec{b}_{in}
 \,=\,\vec{b}_i \,+\,\h \vec{r}_i \,=\,\vec{b}_n \,-\,\h \vec{r}_i$.
  
  \eq{PC1} is a typical cascade equation in which the first term
 describes the reduction   of  the probability to find $n$ dipoles
 due to the possibility that one of $n$ dipoles can decay into two dipoles 
of
 arbitrary sizes  
  , while the second term,  describes  the growth due to the 
splitting
 of $(n-1)$ dipoles into $n$ dipoles.

The initial condition for the DIS scattering is
\be \label{PCIC}
P_1 \Lb Y  =  0,  \vec{r},\vec{b} ; \vec{r}_1,\vec{b}_1\Rb\,\, =\,\,\,\delta^{(2)}\Lb \vec{r}\,-\,\vec{r}_1\Rb\,\delta^{(2)}\Lb \vec{b}\,-\,\vec{b}_1\Rb;~~~~~~~P_{n>1}\Lb Y  = 0; \{r_i\}\Rb \,=\,0
\ee
which corresponds to the fact that we are discussing a dipole of 
 definite size which develops the parton cascade.

Since $P_n\Lb Y ; \{r_i\}\Rb$ is the probability to find dipoles $\{r_i\}$,
 we have the following sum rule 

\be \label{SUMRU}
\sum_{n=1}^\infty\,\int \prod^n_{i=1} d^2 r_i \,d^2 b_i \,P_n\Lb Y ; \{\vec{r}_i\,\vec{b}_i\}\Rb\,\,=\,\,1 ,
\ee
i.e. the sum of all probabilities is equal to 1.

 This QCD  cascade leads to Balitsky-Kovchegov (BK) equation 
\cite{B,K,KOLEB} for the  amplitude and gives the theoretical
 description of the DIS.  We introduce the generating functional\cite{MUCD}

\be \label{Z}
Z\Lb Y, \vec{r},\vec{b}; [u_i]\Rb\,\,=\,\,\sum^{\infty}_{n=1}\int P_n\Lb Y,\vec{r},\vec{b};\{\vec{r}_i\,\vec{b}_i\}\Rb \prod^{n}_{i=1} u\Lb \vec{r}_i\,\vec{b}_i\Rb\,d^2 r_i\,d^2 b_i
\ee
 where $u\Lb \vec{r}_i\,\vec{b}_i\Rb \equiv\,=u_i$ is an arbitrary function.
 The initial conditions of \eq{PCIC}  and the sum rules of \eq{SUMRU} take 
to following form for the functional $Z$:
\begin{subequations}
\bea
Z\Lb Y=0, \vec{r},\vec{b}; [u_i]\Rb &\,\,=\,\,&u\Lb \vec{r},\vec{b}\Rb;\label{ZIC}\\
Z\Lb Y, r,[u_i=1]\Rb &=& 1; \label{ZSR}
\eea
\end{subequations}

Multiplying both parts of \eq{PC1} by $\prod^{n}_{i=1} u\Lb \vec{r}_i\,\vec{b}_i\Rb$ and integrating over $r_i$ and $b_i$ we obtain the following linear functional equation\cite{LELU2};
\begin{subequations}
\bea
&&\hspace{-0.7cm}\frac{\partial Z\Lb Y, \vec{r},\vec{b}; [u_i]\Rb}{\partial \,Y} =\int d^2 r'\,  K\Lb \vec{r}',\vec{r} - \vec{r'}|\vec{r}\Rb\Bigg( - u\Lb r, b\Rb\,\,+\,\,u\Lb \vec{r}',\vec{b} + \h(\vec{r} - \vec{r}') \Rb \,u\Lb \vec{r} - \vec{r}',\vec{b} - \h\vec{r}'\Rb\Bigg) \frac{\delta\,Z}{\delta \,u\Lb r, b \Rb};\label{EQZ}\\
&&  K\Lb \vec{r}',\vec{r} - \vec{r'}|\vec{r}\Rb\,=\frac{\bas}{2 \,\pi}\frac{r^2}{r'^2\,(\vec{r} - \vec{r}')^2} ;\,~~~~~
 \omega_G\Lb r\Rb\,\,=\,\,\int d^2 r'  K\Lb \vec{r}',\vec{r} - \vec{r'}|\vec{r}\Rb; \label{OMG}
 \eea
\end{subequations}
Searching for the solution of the form     $Z\Lb [ u(r_i,b_i,Y)]\Rb$  
for the initial conditions of \eq{ZIC}, \eq{EQZ} can be re-written as
 the non-linear equation \cite{MUCD}:
\be \label{NEQZ}
\frac{\partial Z\Lb Y, \vec{r},\vec{b}; [u_i]\Rb}{\partial \,Y}\,=\,\int d^2 r' K\Lb \vec{r}',\vec{r} - \vec{r'}|\vec{r}\Rb\Bigg\{Z\Lb r',  \vec{b} + \h(\vec{r} - \vec{r}'); [u_i]\Rb  \,Z\Lb \vec{r} - \vec{r'},  \vec{b} - \h\vec{r}'; [u_i]\Rb
\,\,-\,\,Z\Lb Y, \vec{r},\vec{b}; [u_i]\Rb\Bigg\}
\ee
Therefore, the QCD parton cascade of \eq{PC1} takes into account 
 non-linear evolution. However, to obtain the BK equation for the 
scattering amplitude we need to introduce the scattering amplitude
  $\gamma\Lb r_i, b\Rb$, for the interaction of the dipole with the
 target at low energies. Using these amplitudes we can obtain the
 non-linear BK equation from \eq{NEQZ}, since \cite{K}
\be \label{NEXP}
N\Lb Y, r, b\Rb\,=\,\sum^{\infty}_{n=1}\frac{(-1)^{n-1}}{n!}\int \prod^n_{i=1} \Bigg( d^2 r_i \gamma\Lb r_i, b\Rb \frac{\delta}{\delta u_i} \Bigg)Z\Lb Y,r,b,[u_i]\Rb|_{u_i=1}
\ee
Using \eq{EQZ} and \eq{NEXP} we derive the BK equation in the standard form:
\bea 
\frac{\partial}{\partial Y} N\Lb \vec{r}, \vec{b} ,  Y \Rb &=&\int d^2 r'\,K\Lb \vec{r}', \vec{r} - \vec{r}'| \vec{r}\Rb \Bigg\{N\Lb \vec{r}',\vec{b} - \h \Lb \vec{r} - \vec{r}' \Rb, Y\Rb + 
N\Lb\vec{r} - \vec{r}', \vec{b} - \h \vec{r}', Y\Rb \,\,- \,\,N\Lb \vec{r},\vec{b},Y \Rb\nn\\
& & ~~~~~~~- N\Lb\vec{r} - \vec{r}', \vec{b} - \h \vec{r}', Y\Rb   \, N\Lb \vec{r}',\vec{b} - \h \Lb \vec{r} - \vec{r}' \Rb, Y\Rb\Bigg\}\label{NEEQa}
\eea
\subsubsection{Several first iterations}
Our goal is to find the solution to \eq{PC1}. In particular, for the
 multiplicity distribution and for the entropy we wish to find
\be\label{PTILD}
\widetilde{P}_n\Lb Y, r\Rb\,\,=\,\,\int P_n\Lb Y, \vec{r}, \vec{b} ,\{\vec{r}_i,\vec{b}' \Rb  \prod^n_{n=1} d^2 r_i\,d^2 b'
\ee
$\widetilde{P}_n$ is the probability to find $n$ dipoles of all possible
 sizes at the same  values of the impact parameters and, being such, it
 gives $\sigma_n/\sigma_{in}$, which is the multiplicity distribution in
 the QCD parton cascade. The initial and boundary conditions for
 $\widetilde{P}_n\Lb Y, r\Rb$ follows from \eq{PCIC} and \eq{SUMRU}
 and take the form:
\be\label{PTILDIC}
\widetilde{P}_1\Lb Y=0, r, b \Rb\,\,=\,\,1; ~~~\widetilde{P}_n\Lb Y=0, r\Rb\,=\,0\,~\mbox{for}~n>1;
~~~~~\sum^{\infty}_{n=1} \widetilde{P}_1\Lb Y, r, b \Rb\,\,=\,\,1;
\ee

First, let us find  $\widetilde{P}_1\Lb Y, r\Rb$. The equation for $P_1$ has
 the form:

\be \label{P1}
\frac{ \partial P_1\Lb Y, r, b,r_1,b_1\Rb}{\partial \,Y}\,\,=\,\,-\,\omega_G\Lb r_1\Rb P_1\Lb Y, r, b,r_1,b_1\Rb
\ee 
with the initial condition
\be \label{P1IC}
P_1\Lb Y=0, r, b,r_1,b_1\Rb\,\,=\,\,\delta^{(2)}\Lb \vec{r} - \vec{r}_1\Rb\,\delta^{(2)}\Lb \vec{b} - \vec{b}_1\Rb\ee

Therefore for  $\widetilde{P}_1\Lb Y, r, b \Rb$ the equation takes the
 following form:
\be \label{PTILD1}
\frac{\partial \widetilde{P}_1\Lb Y, r, b\Rb}{\partial\,Y}\,\,=\,\,-\,\omega_G\Lb r\Rb  \widetilde{P}_1\Lb Y, r, b\Rb\ee 
with the solution:
\be \label{PTILD11}
\widetilde{P}_1\Lb Y, r, b\Rb\,\,=\,\,e^{-\,\omega_G\Lb r\Rb  \,Y}\ee

The equation for $P_2\Lb Y, r, b,r_1,b', r_2,b'\Rb$ has the following form:

\be \label{P2}
\frac{ \partial P_2\Lb Y, r, b; r_1,b', r_2,b'\Rb}{\partial \,Y}\,\,=
\,\,-\,\Lb \omega_G\Lb r_1\Rb\,+ \omega_G\Lb r_2\Rb\Rb\, P_2\Lb Y, r, b,r_1,b', r_2, b\Rb \,+\,\frac{\bas}{2 \pi}\frac{\Lb \vec{r}_1 + \vec{r}_2\Rb^2}{r^2_1\,r^2_2} P_1\Lb Y, r, b;  \vec{r}_1 + \vec{r_2},b'\Rb
\ee

First, let us estimate the value of $\omega_G\Lb r\Rb$ which is given by
 \eq{OMG}:
\bea \label{OMG1}
\omega_G\Lb r\Rb\,\,&=&\,\,\frac{\bas}{2 \pi}\int d^2 r'\frac{r^2}{r'^2\,\Lb \vec{r} - \vec{r}'\Rb^2} \,=\,
\frac{\bas}{ \pi}\int d^2 r'\frac{r^2}{r'^2\,\Lb r'^2 + \Lb  \vec{r} - \vec{r}'\Rb^2\Rb}\,\,=\,\,\Bigg\{ \int^r_{r_0} \,+\,\int^{\infty}_{r} \Bigg\}  d^2 r'\frac{r^2}{r'^2\,\Lb r'^2 + \Lb  \vec{r} - \vec{r}'\Rb^2\Rb}\nn\\
&\,\,=&\,\,\bas \underbrace{\ln\Lb r^2/r^2_0\Rb}_{r' \,\leq\,r} \,+\,\underbrace{0}_{r' \,\geq\,r}\,\,=\,\,\bas \int^{r^2}_{r^2_0} \frac{d r'^2}{r'^2}
\eea
Hence, only dipoles of size smaller than $r$, contribute to the
 value of $\omega_G\Lb r\Rb$.

We suggest that the solution to \eq{P2} has the following form:

\be \label{SOLP2}
 \int d^2 b'\,P_2\Lb Y, r, b; r_1,b', r_2,b'\Rb\,\,=\,\,\frac{1}{r^2_1\,r^2_2}\,\Theta\Lb r \,- \,r_1\Rb \Theta\Lb r \,- \,r_2\Rb\,p_2\Lb r, b\Rb
 \ee
 where $\Theta\Lb z\Rb$ denotes the step function:$\Theta\Lb z\Rb = 1$ for
 $z > 0$, and $\Theta\Lb z\Rb = 0$ for $z < 0$.
 For the solution of \eq{SOLP2} we can obtain the equation for
 $\widetilde{P}_2\Lb Y, r, b\Rb$, integrating both parts of \eq{P2}
 over $b'$,$r_1$ and $r_2$. It has the form:
 \be \label{PTILD2}
\frac{ \partial \widetilde{P}_2\Lb Y, r\Rb}{\partial \,Y}\,\,=
\,\,-\,2\,\omega_G\Lb r\Rb\, \widetilde{P}_2\Lb Y, r\Rb \,\,+\,\,\omega_G\Lb r\Rb
 \widetilde{P}_1\Lb Y, r\Rb
\ee 
 Using \eq{PTILD11} we obtain that
  \be \label{PTILD22} 
  \widetilde{P}_2\Lb Y, r\Rb\,\,=\,\,e^{ - \,  \omega_G\Lb r\Rb\,Y} \Lb 1 \,-\,e^{ - \,  \omega_G\Lb r\Rb\,Y}\Rb
\ee  
    
    One can see that \eq{PTILD22}   gives   $ \widetilde{P}_2\Lb Y = 0 , r\Rb 
 \,=\,0$ in accord  with \eq{PTILDIC}.
 For small $\omega_G\Lb r \Rb \,Y \,\ll\,\,1$ in the parton cascade only
 two terms  exist: $\widetilde{P}_1\Lb Y,r\Rb$ and $\widetilde{P}_2
\Lb Y,r\Rb$, and \eq{SUMRU}  reduces to
    
    \be \label{SUMRU12}
  \widetilde{P}_1\Lb Y,r\Rb\,+\,\widetilde{P}_2\Lb Y,r\Rb \,\,   \,\xrightarrow{\omega_G\Lb r \Rb \,Y \,\ll\,\,1}\,\,\,\,\underbrace{1\,\,-\,\,    \omega_G\Lb r \Rb \,Y }_{  \widetilde{P}_1}\,\,+\,\, \underbrace{  \omega_G\Lb r \Rb \,Y}_{ \widetilde{P}_2}\,\,=\,\,1
  \ee   
   
   \eq{SUMRU12} shows that  $P_n\Lb Y, r, b, \{ r_i,b'\} \Rb$  are 
negligibly small   
for dipoles with large sizes ( $r_i > r$).

\subsubsection{Solution}
     We suggest to look for the general solution in the form:
   
\be \label{SP}
 P_n\Lb Y, \vec{r }, \vec{b};\,\{\vec{r}_i,\vec{b}'\}\Rb\,\,=\,\,\prod^n_{i=1}\Theta\Lb r - r_i\Rb \frac{1}{r^2_i} \,p_n\Lb Y, r, \Rb
\ee

For such  a solution we can obtain from \eq{PC1} the following equations
 for $\widetilde{P}_n\Lb Y, r\Rb$:
 \be \label{PTILDN}
\frac{ \partial \widetilde{P}_n\Lb Y, r\Rb}{\partial \,Y}\,\,=
\,\,-\,n\,\omega_G\Lb r\Rb\, \widetilde{P}_n\Lb Y, r\Rb \,\,+\,\,\Lb n\,-\,1\Rb\omega_G\Lb r\Rb
 \widetilde{P}_{n - 1}\Lb Y, r\Rb
\ee 
Introducing the Laplace transform:

\be \label{LT}
\widetilde{P}_n\Lb Y, r\Rb\,\,=\,\,\int^{\epsilon\, +\, i \infty}_{\epsilon\, -\, i \infty} \frac{d \omega}{2 \,\pi\,i} \,\widetilde{p}_n\Lb \omega, r\Rb
\ee
we re-write \eq{PTILDN} in the form:
\be \label{PTILDNOM}
\frac{ \partial \widetilde{p}_n\Lb \omega, r\Rb}{\partial \,Y}\,\,=
\,\,-\,n\,\omega_G\Lb r\Rb\, \widetilde{p}_n\Lb , r\Rb \,\,+\,\,\Lb n\,-\,1\Rb\omega_G\Lb r\Rb
 \widetilde{p}_{n - 1}\Lb \omega, r\Rb
\ee 
\eq{PTILDNOM} has the solution:

\be \label{SOLPTIDN}
 \widetilde{p}_n\Lb \omega, r\Rb\,\,=\,\,\Lb n - 1\Rb! \prod^n_{m = 1}\frac{1}{\omega\,\,+\,\,m\,\omega_G\Lb r\Rb}
 \ee

Taking the inverse Laplace transform of function
 $e^{ - \omega_G\Lb r \Rb\,Y}\Lb 1 \,\,-\,\,e^{ - \omega_G\Lb r \Rb\,Y}\Rb^{n - 1}$, we have
\be \label{SOLPN}
\int^{\infty}_{0}\,d Y \,e^{- \omega\,Y} e^{ - \omega_G\Lb r \Rb\,Y}\Lb 1 \,\,-\,\,e^{ - \omega_G\Lb r \Rb\,Y}\Rb^{n - 1}\,\,=\,\,\Lb n - 1\Rb! \prod^n_{m = 1}\frac{\omega_G\Lb r\Rb}{\omega\,\,+\,\,m\,\omega_G\Lb r\Rb}\,\,=\,\, \widetilde{p}_n\Lb \omega, r\Rb
\ee
Hence we have the following solution:
\be \label{SOLPN1}
\widetilde{P}_n\Lb Y, r\Rb\,\,=\,\,e^{ - \omega_G\Lb r \Rb\,Y}\Lb 1 \,\,-\,\,e^{ - \omega_G\Lb r \Rb\,Y}\Rb^{n - 1}
\ee
It is easy to see that \eq{SOLPN1} satisfies the initial conditions and
 the sum rules of \eq{PTILDIC}.

 \subsection{Multiplicity distribution and entropy of the parton cascade}
As has been mentioned 
\be \label{XN}
\frac{\sigma_n}{\sigma_{ \rm in}}\,\,=\,\,\widetilde{P}_n\Lb Y, r\Rb\,
\ee
and, therefore, determines the multiplicity distribution. Calculating 
the average $N$
\be \label{AVN}
N\,\,=\,\,\sum^{\infty}_{n=1} n \,\frac{\sigma_n}{\sigma_{ \rm in}}\,\,=\,\,\sum^{\infty}_{n=1} \,n\, 
\widetilde{P}_n\Lb Y, r\Rb\,\,=\,\,e^{\omega\Lb  r\Rb\,Y}
\ee
we see that  this distribution can be written in the form:
\be \label{NDIST}
\frac{\sigma_n}{\sigma_{ \rm in}}\,\,=\,\,\frac{1}{N}\,\Lb \frac{N\,-\,1}{N}\Rb^{n - 1}\,=\,\frac{1}{\bar{N}}\Lb \frac{\bar{N}}{\bar{N} + 1}\Rb^n ,
\ee
where we have denoted $\bar{N} = N - 1$.

We can compare \eq{NDIST} with the general form of the negative binomial
 distribution (NBD) 
\be \label{NBD}
P^{\rm NBD}\Lb r, \bar{n},n\Rb\,\,=\,\,\Lb \frac{r}{r + 
\langle n \rangle}\Rb^r 
\frac{\Gamma\Lb n + r\Rb}{n!\, \Gamma\Lb r \Rb}\Lb 
\frac{\langle n \rangle}{r\,+\,\langle n \rangle}\Rb^n  ,
\ee
One can see that \eq{NDIST} can be re-written as
\be \label{MDNB}
\frac{\sigma_n}{\sigma_{\rm in}}\,\,=\frac{\bar{N} }{\bar{N}\,+\,1}P^{\rm NBD}\Lb 1, \bar{N},n\Rb ,
\ee
where $\sigma_n$ is the cross section for producing $n$ hadrons in a
 collision, and $\sigma_{\rm in}$ is the inelastic cross section. 
Therefore at large $\bar{N}$  our distribution is close to the 
negative binomial distribution with number of failures $r =1$ and
 with probability of success $p = \bar{N}/\Lb \bar{N} + 1\Rb$. \eq{NDIST}
 coincides with the multiplicity distribution in the parton model (see 
below
 and Ref.\cite{KHLE}). The difference is only in the expression  for the
 average multiplicity ($N$).

Having \eq{NDIST} we can calculate 
the von Neumann entropy of the parton cascade (see \eq{I1}),  given
 by the Gibbs formula 

\be \label{GIBBS}
S_{\rm parton \,cascade} ,\,=\,\,- \sum_{n} \,p_n \,\ln p_n 
\ee
where $p_n$ is the probabilities of micro-states. In the parton cascade we
 can  identify $p_n$ with $\widetilde{P}_n\Lb Y, r\Rb$ , reducing
 \eq{GIBBS} to the following expression:
\bea \label{GIBBS}
S_{\rm parton \,cascade} \,\,&=&\,\,- \sum_{n} \widetilde{P}_n\Lb Y, r\Rb \,\ln \widetilde{P}_n\Lb Y, r\Rb
\,=\,\sum_n\Bigg(\ln \bar{N} \,-\,n \,\ln\Lb\frac{\bar{N}}{\bar{N}+1}\Rb\Bigg) \frac{1}{\bar{N}}\Lb \frac{\bar{N}}{\bar{N} + 1}\Rb^n\nn\\
&=&\,\,\ln \bar{N} \,\,+\,\ln\Lb \frac{\bar{N}}{\bar{N} \,+\,1}\Rb\Lb 1 + \frac{1}{\bar{N}}\Rb\,\,\xrightarrow{\bar{N} \,\gg\,1} \ln\Lb N - 1\Rb\,\,=\,\,\omega_G\Lb r \Rb\,Y
\eea

     \begin{figure}[ht]
     \begin{center}
     \includegraphics[width=0.7\textwidth]{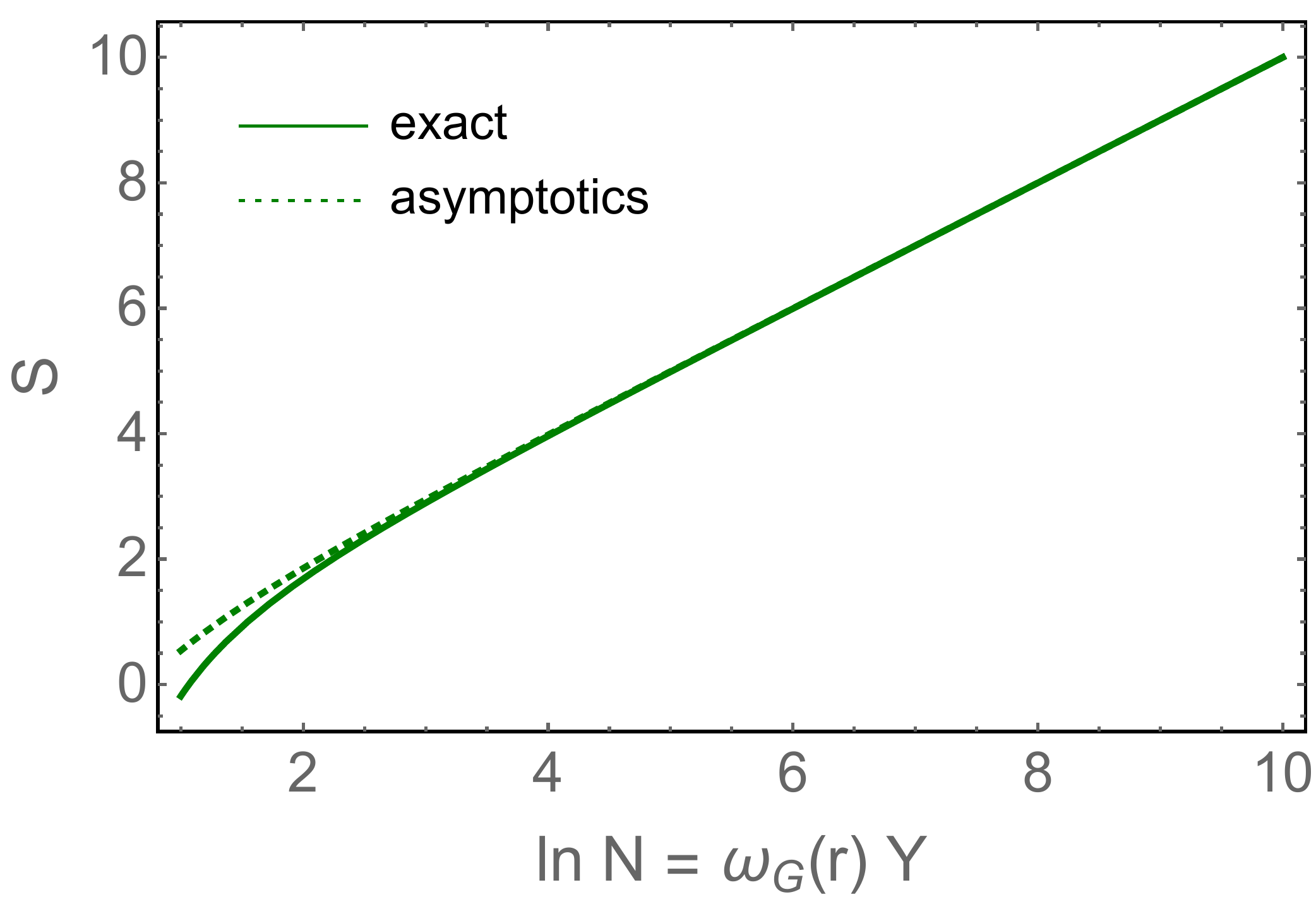} 
     \end{center}    
      \caption{ Entropy $S$ versus $ \ln N \,\,=\,\,\omega_G\Lb r\Rb\,Y$ 
(see \eq{GIBBS}. $S_{\rm asymp}\,\,=\,\,\ln N\,\,=\,\,\omega_G\Lb r \Rb\,Y$.}
\label{s}
   \end{figure}
\eq{GIBBS} shows that at large $Y$ 
 all probabilities $\widetilde{P}_n$  become equal and small, of the
 order of $\widetilde{P}_n \,\,\sim \frac{1}{N}$ . It is well known
 that this equipartitioning of micro-states maximizes the von Neumann
 entropy and describes the maximally entangled state. We thus conclude
 that at large $Y$  the fast dipole  represents a maximally entangled 
quantum state of partons.
 
\subsection{QCD motivated parton model}
The main assumption of the parton model\cite{FEYN,BJP,Gribov} is 
that all partons have average transverse momentum which does not depend
 on energy. Therefore, we can obtain the parton model from the QCD
 cascade assuming that the unknown confinement of gluons leads to 
the QCD cascade for the dipole of fixed size. In this case the
 cascade equation take the following form:
\be \label{EQPM}
 \frac{d P_n\Lb Y\Rb}{d Y}\,\,=\,\,- \Delta\,n\,  P_n\Lb Y\Rb \,\,+\,\,\Lb n - 1 \Rb \Delta P_{n-1}\Lb Y .\Rb
\ee
where  $P_n\Lb Y \Rb$ is the probability to find $n$ dipoles
 (of a fixed size in our model) at  rapidity $Y$ and $ \Delta = \omega_G\Lb
 r = r_0\Rb$.

Using the Laplace transform of \eq{LT} we obtain the solution to \eq{EQPM}
 in the form:
\be \label{SOLPM}
 P_n\Lb \omega\Rb\,\,=\,\,\Lb n - 1\Rb! \prod^n_{m = 1}\frac{1}{\omega\,\,
+\,\,m\,\Delta}
 \ee

Using  \eq{SOLPN}, we see that the solution of \eq{EQPM} has the form:

\be \label{PNPM}
\frac{\sigma_{n}}{\sigma_{in}}\,\,\equiv\,\,P_n\Lb Y \Rb\,\,=\,\,e^{\,-\,\Delta\, Y }\Lb 1\,-\, e^{\,-\,\Delta\,Y}\Rb^{n - 1} \,\,=\,\,\frac{1}{N}\Lb \frac{N - 1}{N}\Rb^{n - 1}
\ee
which is a direct generalization of \eq{SOLPN1}. In \eq{PNPM}, $N$ is the
 average number of the partons, which is equal to $N = \exp\Lb \Delta\,Y\Rb$.
 In the parton model the average number of partons is related to the deep
 inelastic structure function. Therefore, in Ref.\cite{KHLE} it is assumed 
that 
\be \label{XGPM}
N\,\,=\,\,xG\Lb x, Q^2\Rb
\ee
where $x G$ is the gluon structure function. In this case $\Delta$ can
 be identified  with the intercept of the BFKL Pomeron\cite{BFKL}. One can
 see that $N = \exp\Lb \omega_G\Lb r\Rb\,Y\Rb$ this is certainly  not 
the same
 as a solution of the QCD evolution for the gluon structure function. 

 
\subsection{
Mutiplicity distribution for the parton cascade in DIS}


As we have seen (see \eq{SP}) our solution describes the evolution
 in the system of partons with smaller sizes of dipoles than the
 initial fastest dipole. On the other side, only partons of larger
 than initial parton size, contribute to the structure function. In
 double log approximation the emission of such dipoles leads to $xG
 \propto \exp\Lb 2 \sqrt{\bas Y\,\ln\Lb R^2/r^2\Rb}\Rb$, where $R$ is
 the size of the target. We need to consider how the produced dipoles 
interact
 with the target. We measure the gluon structure function in the
 experiment in which $r^2$ of the fastest dipole is about $r^2
 \ sim 1/Q^2 
 \ll\, R^2$.  Since all $r_i \leq \,r$ the $n$ produced dipoles
 interact with the target at rapidity $Y$ and since $r_i\, < \,R$ the
 amplitude of this interaction is proportional to $r_i^2/R^2$. This fact
  completely changes the structure of the cascade. Let us illustrate this,
 considering $P_2\Lb Y, r; r_1,r_2\Rb$. The amplitude of interaction is
 proportional to $\rho_2(Y, r, b; r_1,r_2,b') \,\,\equiv\,\,  (r^2_1 + r^2_2)P_2\Lb Y,r, b;  r_1,r_2, b'\Rb$.  Let us look at \eq{P2} for $r_i \,>\,r$. The term with gluon $\omega_G\Lb r_i\Rb$ does not contribute since as we have discussed only $r_i > r$ contribute in this term. Therefore the equation reduces to the following one:
\be \label{RHO2}
\frac{\partial\Lb \rho_2(Y, r, b; r_1,r_2,b'\Rb}{\partial \,Y}\,\,=\,\,\frac{r^2}{r^4_1} 2\,r^2_1 P\Lb Y,r, b;r_1.b'\Rb=\,
2 \frac{1}{r^2_1} P\Lb Y,r, b;r_1.b'\Rb
\ee

Therefore, we infer that  dipole sizes  larger than  $r$, contribute
 to the scattering amplitude of our interest,   and lead to large 
$\rho_2$.

Generally speaking the scattering amplitude can be written in the
 form\cite{K,LELU2}:
\be \label{N1}
N(Y,\,r,\,b)\,=\,-\,\,\sum^{\infty}_{n=1}\,(-1)^n\,
\rho^p_n(r_1,\, b_1,\,\ldots\,r_n,\,b_n\,;\,Y\,-\,Y_0)
\,\,\prod^n_{i =1}\,N(Y_0,\,r_i,\,b_i)\,\,
d^2\, r_i\, \, d^2\, b_i \,.
\ee

where $N(Y_0,\,r_i,\,b_i)$ is the amplitude of the interaction of dipole
 $r_i$ with the target at low energy $Y=Y_0$, and the $n$-dipole densities
 in the projectile 
$\rho^p_n(r_1, b_1,\ldots\,,r_n, b_n)$
 are defined as follows:
\be \label{N2}
\rho^p_n(r_1, b_1\,
\ldots\,,r_n, b_n; Y\,-\,Y_0)\,=\,\frac{1}{n!}\,\prod^n_{i =1}
\,\frac{\delta}{\delta
u_i } \,Z\left(Y\,-\,Y_0;\,[u] \right)|_{u=1}
\ee
For $\rho_n$ we obtain\cite{LELU2} :
\bea \label{N3}
\frac{\partial \,\rho^p_n(r_1, b_1\,\ldots\,,r_n, b_n)}{ 
\bar{\alpha}_s\,\partial\,Y}\,\,&=&\,\
-\,\sum_{i=1}^n
 \,\,\omega(r_i)\,\,\rho^p_n(r_1, b_1\,\ldots\,,r_n, b_n)\,\,+\,\,2\,\sum_{i=1}^n\,
\int\,\frac{d^2\,r'}{2\,\pi}\,
\frac{r'^2}{r^2_i\,(\vec{r}_i\,-\,\vec{r}')^2}\,
\rho^p_n(\ldots\,r', b_i-r'/2\dots)\nn\\
 & & 
\,
+\,\sum_{i=1}^{n-1}\,\frac{(\vec{r}_i + \vec{r}_n)^2}
{(2\,\pi)\,r^2_i\,r^2_n}\,
\rho^p_{n-1}(\ldots\,(\vec{r}_i\,+\,\vec{r}_n), b_{in}\dots).
\eea
For $\rho_1$ 
we have  the linear equation: 
\be \label{N4}
\frac{\partial \,\rho^p_1(Y; r_1, b)}{ 
\bas\,\partial\,Y}\,\,=\,\,-\,\,\omega_G\Lb r_1\Rb\rho^p_1(Y; r_1, b) \,\,+\,2\,\int\,\frac{d^2\,r'}{2\,\pi}\,
\frac{r'^2}{r^2_1\,(\vec{r}_1\,-\,\vec{r}')^2}\,
\bar{\rho}^p_1\Lb Y, r',b\Rb
\ee
Introducing $\bar{\rho}^p_1\Lb Y;  r_1,b\Rb\,\,=\,\,r^2_1\,\rho^p_1 \Lb
 Y, r',b\Rb$, we obtain for $\bar{\rho}^p_1\Lb Y  r_1,b\Rb$ the BFKL equation:
\be \label{N5}
\frac{\partial \,\bar{\rho}^p_1(Y; r_1, b)}{ 
\bas\,\partial\,Y}\,\,=\,\,-\,\,\omega_G\Lb r_1\Rb\bar{\rho}^p_1(Y; r_1, b) \,\,+\,2\,\int\,\frac{d^2\,r'}{2\,\pi}\,
\frac{1}{(\vec{r}_1\,-\,\vec{r}')^2}\,
\bar{\rho}^p_1\Lb Y, r',b\Rb
\ee
The physical meaning of $\rho^p_1$ is clear from \eq{N2}: it is the mean
 number of dipoles with size $r_1$  that have been produced. The
 multiplicity, which we needed in DIS,  is the number of dipoles
 with sizes larger than $r \sim 1/Q$. It is equal to
\be \label{N6}
N^p_1\Lb Y, r \Rb\,\,=\,\,\,\int_{r} d^2 r_1, d^2 b \rho^p_1(Y; r_1, b)\,\,=\,\,\int^\xi d \xi'  d^2 b \bar{\rho}^p_1\Lb Y, \xi',b\Rb\,\,=\,\,\Big{<} n\Big{>}
\ee
where $\xi \,=\,\ln (1/r^2)$.

In the  double log approximation (DLA) of perturbative QCD, \eq{N5} can be 
re-written in the form
\be \label{N7}
\frac{ \partial^2\,N^p_1\Lb Y, r\Rb}{\bas\,\partial \,Y\,\partial\,\xi}\,\,=\,\,N^p_1\Lb Y, r\Rb
\ee
It is worthwhile mentioning that $N^p_1$ is the gluon structure
 function in the DLA.

Equation for $\bar{\rho}^p_2\Lb Y ; r_1, r_2, b\Rb$ has the form\footnote{For
 simplicity of presentation we took $b\,\gg\,r_i$ .}:

\bea \label{N8}
\frac{\partial \,\rho^p_2(Y; r_1, r_2, b)}{ 
\bas\,\partial\,Y}\,\,&=&\,\,-\,\,\Lb \omega_G\Lb r_1\Rb\,+\,\omega_G\Lb r_2\Rb\Rb \rho^p_2(Y; r_1, r_2 b) \,\,+\,2\,\int\,\frac{d^2\,r'}{2\,\pi}\,
\frac{r'^2}{r^2_1\,(\vec{r}_1\,-\,\vec{r}')^2}\,
\rho^p_2\Lb Y, r',b,r_2,b\Rb\nn\\
&\,\,+&\,\,2\,\int\,\frac{d^2\,r'}{2\,\pi}\,
\frac{r'^2}{r^2_2\,(\vec{r}_2\,-\,\vec{r}')^2}\,
\rho^p_2\Lb Y, r_1, r', b\Rb\,\,+\,\, \frac{(\vec{r}_1 + \vec{r}_2)^2}
{(2\,\pi)\,r^2_1\,r^2_2}\,\rho^p_{1}\Lb Y; \vec{r}_1\,+\,\vec{r}_2), b\Rb
\eea

However, to find the multiplicity distribution we need to introduce moments
 (see \eq{PTILD}):
\be \label{N7}
N^p_n\Lb Y,  r\Rb\,\,=\,\,\int  \prod^n_{i=1} d^2 r_i  d^2 b \,\rho^p_n\Lb  Y,  \{ r_i\},b\Rb\,\,=\,\,\int  \prod^n_{i=1}\frac{ d^2 r_i }{r^2_i }  d^2 b\, \bar{\rho}^p_n\Lb Y, \{r_i\},b\Rb
\ee

$N^p_2$ is equal to
\be \label{N8}
N^p_2\Lb Y, r\Rb\,\,=\,\,\int^\xi d \xi_1 \int^\xi d \xi_2 \,\int d^2 b\,\, \bar{\rho}^p_2\Lb Y, \xi_1,\xi_2,b\Rb
\ee
which gives $ \Big{<} \frac{ n (n - 1)}{2}\Big{>}$.

Equation for $\rho^p_2$ can be re-written in the following form for
 $\bar{\rho}^p_2$:
\bea \label{N9}
\frac{\partial \,\bar{\rho}^p_2(Y; r_1, r_2, b)}{ 
\bas\,\partial\,Y}\,\,&=&\,\,-\,\,\Lb \omega_G\Lb r_1\Rb\,+\,\omega_G\Lb r_2\Rb\Rb\bar{ \rho}^p_2(Y; r_1, r_2 b) \,\,+\,2\,\int\,\frac{d^2\,r'}{2\,\pi}\,
\frac{1}{(\vec{r}_1\,-\,\vec{r}')^2}\,
\bar{\rho}^p_2\Lb Y, r',b,r_2,b\Rb\nn\\
&\,\,+&\,\,2\,\int\,\frac{d^2\,r'}{2\,\pi}\,
\frac{1}{(\vec{r}_2\,-\,\vec{r}')^2}\,
\bar{\rho}^p_2\Lb Y, r_1, r', b\Rb\,\,+\,\,\bar{\rho}^p_{1}\Lb Y; \vec{r}_1\,+\,\vec{r}_2, b\Rb
\eea
or in DLA it takes the form:
\be \label{N10}
\frac{\partial \,\bar{\rho}^p_2(Y; \xi_1,\xi_2, b)}{ 
\bas\,\partial\,Y} \,\,=\,\,\,\int^{\xi_1} \!d \xi' \,
\bar{\rho}^p_2\Lb Y,\xi', \xi_2,b \Rb\,\,+\,\,\int^{\xi_2} \!d \xi' \,
\bar{\rho}^p_2\Lb Y,\xi_1, \xi',b \Rb\,\,+\,\,\bar{\rho}^p_1\Lb \xi_1 \approx \xi_2\Rb
\ee
Note, that the gluon reggeization does not contribute in DLA, since it
 describes the contribution of  distances $r_i \,<\,r$ (see discussion
 above). Integrating \eq{N10} over $\xi_1$ and $\xi_2$ we obtain:
\be \label{N11}
\frac{\partial N^p_2\Lb Y, \xi\Rb}{ 
\bas\,\partial\,Y} \,\,=\,\,\,2\,\int^{\xi}\!d \xi' \,N^p_2\Lb Y, \xi' \Rb\,\,+\,\,\int^{\xi}\!d \xi' \,N^p_1\Lb Y, \xi'\Rb
\ee

The general solution to \eq{N11} has a form: $N^p_2\Lb Y, \xi\Rb\,\,=\,\, N^{p,\mbox{\tiny homog}}_2\Lb Y, \xi\Rb\,  \,\,-\,\,  N^p_1\Lb Y, \xi\Rb$, where $N^{p,\mbox{\tiny homog}}_2\Lb Y, \xi\Rb$ is the solution of the homogenous equation:
\be \label{N12}
\frac{\partial N^{p,\mbox{\tiny homog}}_2\Lb Y, \xi\Rb}{ 
\bas\,\partial\,Y} \,\,=\,\,\,2\,\int^{\xi}\!d \xi' N^{p,\mbox{\tiny homog}}_2\Lb Y, \xi\Rb
\ee
The solution of \eq{N12} has the form:

\be \label{N13}
N^{p,\mbox{\tiny homog}}_2\Lb Y, \xi\Rb\,\,\,=\,\,\,\int^{\epsilon \,+\,i\,\infty}_{\epsilon \,-\,i\,\infty} \frac{d \gamma}{2\,\pi\,i}\,e^{ \frac{2\,\bas}{\gamma}Y \,\,+\,\,\gamma\,\xi}\,n_{in}\Lb \gamma\Rb
\ee
where $ \omega(\gamma) \,=\,\frac{\bas}{\gamma}$ is the DLA limit of
 the BFKL kernel:

\be \label{BFKLKER}
\omega\Lb  \gamma\Rb\,\,=\,\,\bas\,\chi\Lb \gamma \Rb\,\,\,=\,\,\,\bas \Lb 2 \psi\Lb 1\Rb \,-\,\psi\Lb \gamma\Rb\,-\,\psi\Lb 1 - \gamma\Rb\Rb
\ee
where $\psi(z)$ is Euler gamma function (see \cite{RY} formula{ \bf 8.36}).

We select $n_{in}\Lb \gamma\Rb \,\,=\,\,1/\gamma$, since at $Y=0$  we have
 only one dipole and $N_2\,=\, < n (n - 1)/2>  = 0$. Taking the integral
 over $\gamma$ using the method of steepest descent, we obtain
\be \label{N14}
N^{p,\mbox{\tiny homog}}_2\Lb Y, \xi\Rb\,\,=\,\,\Lb \frac{\pi}{2 \sqrt{2\,\bas\,Y\,\xi}}\Rb^{1/2} \exp\Lb 2 \sqrt{2 \bas\,Y\,\xi}\Rb
\ee

First we wish to note that \eq{N14} leads to   $N^{p,\mbox{\tiny homog}}_2\Lb Y, \xi\Rb\, \,\neq\,\,\Lb  N^p_1\Lb Y, \xi\Rb\Rb^2$. However, in the diffusion approximation for the BFKL kernel 
\bea \label{DIFKER}
&&\omega\Lb  \gamma\Rb\,\,=\,\,\Delta_{\rm BFKL} \,+\,D\Lb 1/2 - \gamma\Rb^2\,\,=\,\,\Delta_{\rm BFKL} \,-\,D\,\nu^2~~~\mbox{with}~~\gamma\,=\,\h\, +\, i\,\nu;\\
&&\Delta_{\rm BFKL} \,=\,4\ln 2 \bas;\,\,D\,\,=\,\,14 \zeta(3) \bas \,=\,16.828\,\bas;\nn
\eea
 the main contribution stems from 
$\omega =  \Delta_{\rm BFKL}$ and $N^{p,\mbox{\tiny homog}}_2\Lb Y, \xi\Rb\,
\,=\,\,\Lb  N^p_1\Lb Y, \xi\Rb\Rb^2$,
if we neglect the contributions at $\gamma \neq 1/2$ $\nu \,\neq\,0$). 
 Taking the integral over $\gamma$ using the method of steepest descent 
for
 the kernel of \eq{DIFKER}, one can see that the values of the saddle
 point for $\nu$ are equal 
\be \label{SPNU}
\nu_{\rm SP} \,\,=\,\,\frac{\xi}{2\,D\,n\,Y}
\ee 
for $N^P_n$. Therefore, for large $Y$ as well as for large $n$ we, indeed,
 can consider $\nu_{\rm SP} \,\to\,0$.
 For this special case we  have

 \be \label{N15}
  N^p_2\Lb Y, \xi\Rb\,\,=\,\,\Lb  N^p_1\Lb Y, \xi\Rb\Rb^2\,\,-\,\, N^p_1\Lb Y, \xi\Rb
  \ee
  
Comparing with the multiplicity distribution of \eq{NDIST}, one can see
 that \eq{N12} gives  the factorial moment $\Big{<}\frac{n (n - 1)}{2}\Big{>}$
 of this distribution with $ < n > = N^p_1$.

For $N^p_n$, the equation follows from \eq{N8} which in DLA takes the 
form:
\be \label{N16}
\frac{\partial N^p_n\Lb Y, \xi\Rb}{ 
\bas\,\partial\,Y} \,\,=\,\,\,n \,\int^{\xi}\!d \xi' \,N^p_n\Lb Y, \xi' \Rb\,\,+\,\,\Lb n - 1\Rb\,\int^{\xi}\!d \xi' \,N^p_{n-1}\Lb Y, \xi'\Rb
\ee
The solution to this equation has the form:
\be \label{N17}
N^{p}_n\Lb Y, \xi\Rb\,\,\,=\,\,\,\int^{\epsilon \,+\,i\,\infty}_{\epsilon \,-\,i\,\infty} \frac{d \gamma}{2\,\pi\,i}\,\frac{1}{\gamma}e^{ \omega\Lb \gamma\Rb\,Y \,\,+\,\,\gamma\,\xi}\, \Big\{e^{  \omega\Lb \gamma\Rb\,Y }\,\,\,-\,\,1\Big\}^{n - 1} \ee
with $\omega\Lb \gamma\Rb = \bas/\gamma$ in DLA.

Comparing \eq{N17} with the moments  $ M_q\,=\,< \frac{n!}{q! (n - q)!}>$ for
 the multiplicity distribution of \eq{NDIST}:
\be\label{MQ}
N_1\,\,=\,\,\bar{n}; ~~~N_k(k>1)\,\,=\,\,\Bigg{<} \frac{n!}{q! (n - q)!}\Bigg{>}\,\,=\,\,\bar{n} \Lb \bar{n}\,-\,1\Rb^{k - 1}
\ee
one can see that  $N^{p}_n\Lb Y, \xi\Rb$  coincide  with these moments only 
 if we take into account  the main exponential behaviour at
 $\gamma = 1/2$. As we have seen above, for large $n$ the value
 of the saddle point for $\nu$ (see \eq{SPNU} ) indeed approaching zero.

Hence, we infer that the QCD parton cascade in DIS leads to
 the multiplicity distribution of \eq{NDIST} with 
$N = xG\Lb Q,x\Rb$ at $ x\,\to\,0$, as  is expected from the
 parton model of section II-E, but    $xG$   should satisfy the BFKL
 evolution equation, and the  accuracy  of  \eq{NDIST}  is not
 very  precise  at  small $n$.

~

\section{Multiplicity distribution in Colour Glass Condensate (CGC) approach}


In Ref.\cite{KOV3} the density matrix is calculated in the CGC
 approach, using  the CGC wave function from Refs.\cite{KLW,AKLP}.
 In CGC approach the 
 large fraction of
momentum is carried by the valence quarks and gluons. These fast 
partons emit low energy gluons whose lifetime is much shorter
  than the valence partons. In  other words the valence ("hard") partons
 can be treated as static sources of the soft gluons. The wave function 
of such a system of partons can be written in the form:
 \be \label{CGC1}
 |\psi>\,\,=\,\,| v> \otimes\, |s>
 \ee
 where $| v >$ characterizes the valence degrees of freedom, while
 $| s >$ denotes the wave function of the soft gluon in the presence of 
the
 valence partons. Sign $\otimes$ does not denote a direct product,
 since  the wave function of a soft gluon depends on the valence degrees
 of freedom. Using that 
 \be \label{CGC2} 
 |s>\,=\,{\cal C}|0>~~~\mbox{with}~~{\cal C} = \exp\Lb 2 i tr\int \frac{d^2 k}{(2 \pi)^2}b^i(k) \phi^a_i(k)\Rb
 \ee
 where $\phi_i(k) \equiv\,a^+_i\,+\,a(-k)$ and $b^i_a = g \rho_a(k)
 \frac{i \vec{k}_i}{k^2} + ..$( $\rho^a$ is the charge density of 
the valence partons), and McLerran-Venugopalan (MV) model for wave
 function $| v >$, the  density matrix:
 \be \label{CGC3}
 \hat{\rho}\,\,=| v> \otimes | s> <s | \otimes<v|
 \ee
 is calculated in Ref.\cite{KOV3}. The result of these calculation
 is the following:
 \be \label{CGC4}
 <l_c(q),m_c(-q)|\hat{\rho}|\alpha_c(q), \beta_c(-q)>\,\,=\,\,\Lb 1 \,-\,R\Rb \frac{(l + \beta)!}{\sqrt{l!\,m!\,\alpha!\,\beta!}}\Lb \frac{R}{2}\Rb^{l\,+\,\beta}\,\delta_{l + \beta,m + \alpha}
 \ee
 with
 
\be \label{CGC5}
R\,\,=\,\,\Lb 1\,+\,\frac{q^2}{2\,g^2\,\mu^2}\Rb^{-1}
\ee
where    $g$ is QCD coupling constant, and    $\mu^2$   determines the 
colour charge density in the valence wave function in the MV model\cite{MV}.

 For the multiplicity distribution we  only need the diagonal elements 
of the density matrix with $l  = \alpha$ and $m = \beta$, and the 
multiplicity is $n\,=\,l\,+\,m$. Plugging these $l$ and $m$ into
 \eq{CGC4} we obtain for the multiplicity distribution
 \be \label{CGC6}
 \frac{\sigma_n}{\sigma_{\rm in}}\,\,=\,\,\Lb 1 \,-\,R\Rb\sum_m \frac{n!}{m!\,(n - m)! }\Lb \frac{R}{2}\Rb^{n}\,\,=\,\,\Lb 1 \,-\,R\Rb\,R^n
 \ee
 Calculating average $n$ = N  we obtain 
  \be \label{CGC7} 
 N\,\,=\,\,\Lb 1\,-\,R\Rb^{-1} 
 \ee
 and the multiplicity distribution can be re-written in the form:
 
 \be \label{CGC8}
  \frac{\sigma_n}{\sigma_{\rm in}}\,\,=\,\,\frac{1}{N} \Lb \frac{N - 1}{N}\Rb^n\,\,\,=\,\,\,\frac{1}{\bar{N}}\Lb \frac{\bar{N}}{\bar{N} + 1}\Rb^n 
  \ee
  
  We   stress that \eq{CGC8} coincides with \eq{NDIST}, 
that we derived for the QCD parton cascade.
  
 \section{Multiplicity distribution in hadron-hadron scattering}
 \subsection{The interaction of two dipoles at high energies}
\begin{figure}[ht]
     \begin{center}
     \begin{tabular}{c c c}
     \includegraphics[width= 0.6\textwidth]{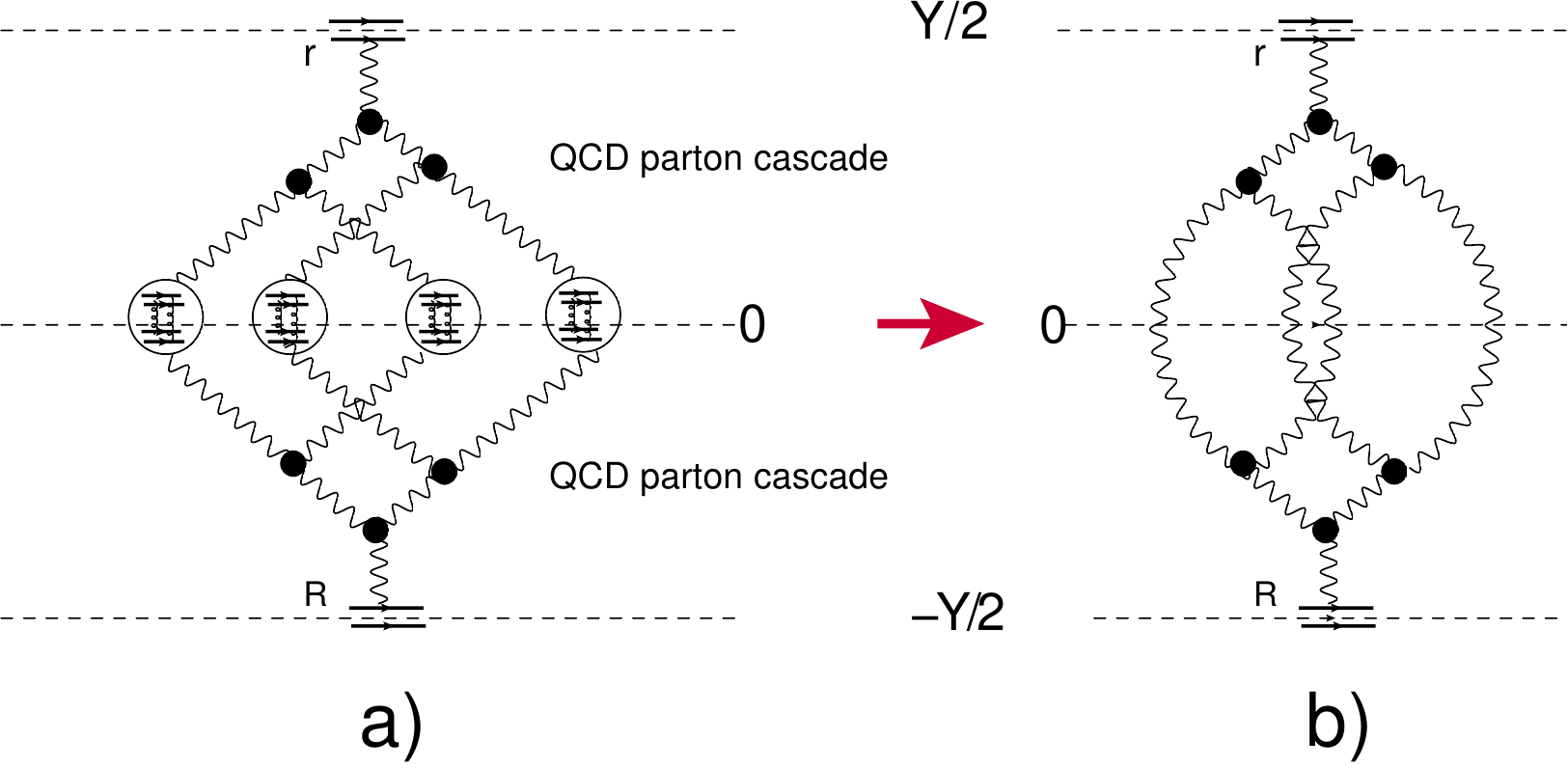} &~~~~~~~~~& \includegraphics[width=0.255\textwidth]{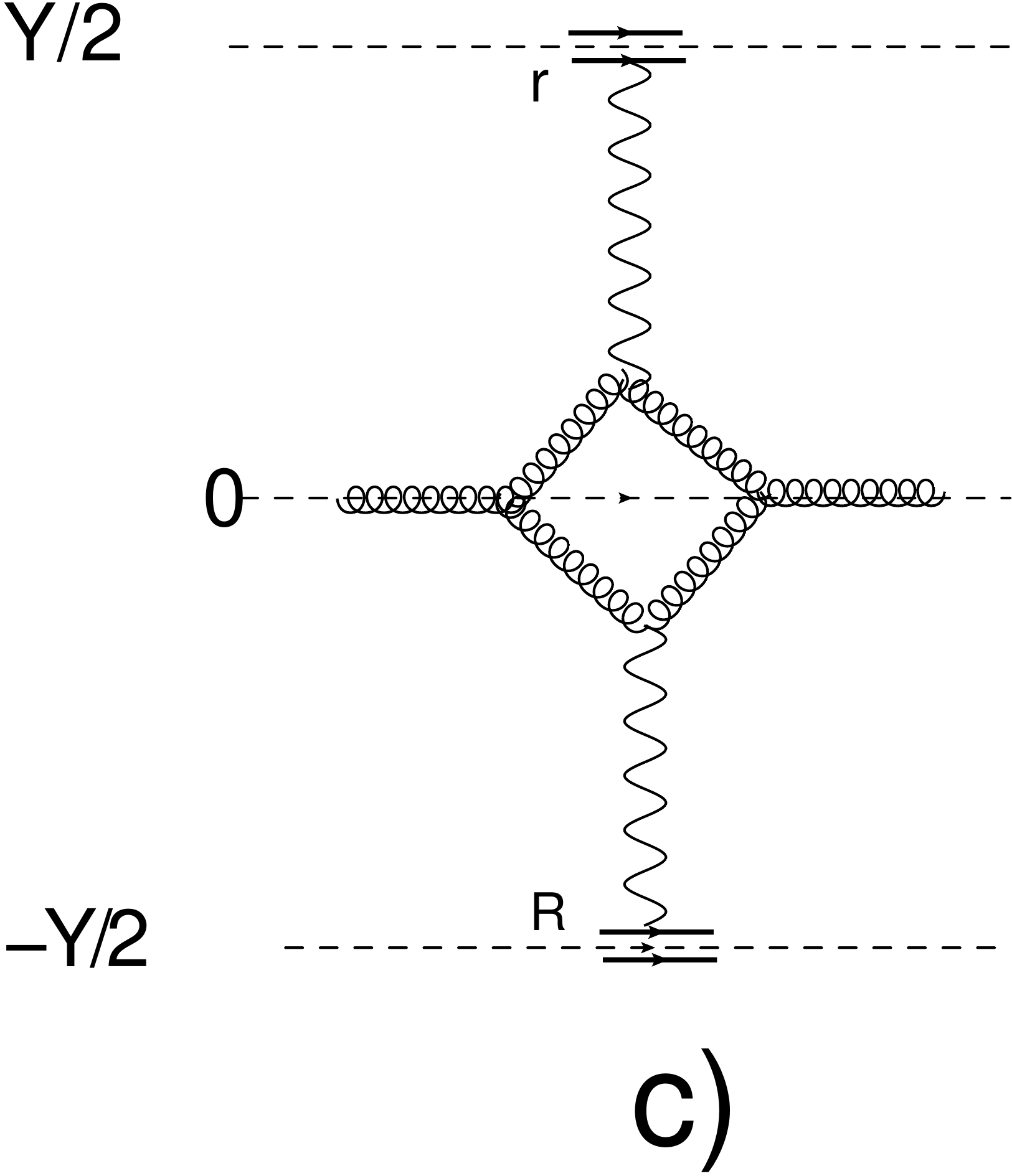}\\
    \end{tabular}     \end{center}    
      \caption{ Scattering amplitude for the interaction of two dipoles
 with sizes:  $r$ and $R$ at high energy in MPSI approach
 (see \fig{mpsi}-a and \fig{mpsi}-b). The amplitudes of
 interaction of two dipoles in the Born approximation of
 perturbative QCD ( $N\Lb r_i,r'_i,b"_i\Rb$ in \eq{HH1})
 are shown   the  white circles. The wavy  lines denote 
the BFKL Pomerons. \fig{mpsi}-c shows the Mueller diagram\cite{MUDIA}
 for inclusive production of gluons. }
\label{mpsi}
   \end{figure}

 We first  consider the high energy interactions of two dipoles
 with sizes $r$ and $R$ and with $r \,\sim\, R$.
 In Ref.\cite{AKLL1} it is shown that in the limited range of rapidities,
 which is given by \eq{I1}, we can safely apply 
 the  Muller, Patel, Salam and Iancu approach for this 
scattering \cite{MUPA,MPSI}(see
 \fig{mpsi}-a).  
The scattering amplitude in this approach can be written in the following
 form\cite{LELU2}:

\bea \label{HH1}
 N\Lb Y, r, R, b \Rb \,\,&=&\,\,-\,\, \sum^{\infty}_{n =1}\,\,(-1)^n\,
\int\,\,\rho^t_n\Lb \vec{r}_1, \,\vec{b}'_1 ,\ldots \,, \vec{r}_n,\,  \vec{b}'_n;\,\h Y\Rb\,\,
\rho^p_n\Lb \vec{r}'_1, \, \vec{b} - \vec{b}'_1 - \vec{b}_1''\,,\ldots\,,\vec{r}_n,\,  \vec{b} - \vec{b}'_n - \vec{b}_n'';\, -\h Y\Rb \nn\\
& \times& \,\prod^n_{i =1}\, \,
d^2\, r_i \,
\,\,\prod^n_{j =1}\, \,d^2 \,r'_j \, d^2 b'_j d^2 b''_j\,N^{\rm BA}\Lb r_i, r'_i,b''_{i}\Rb
\eea

where $\rho^t_n$ and     $\rho^p_n$ are the parton densities in the target 
and projectile, respectively. These densities can be calculated from $P_n$ 
using \eq{N2}. $N^{\rm BA} $ is the scattering amplitude of two dipoles in
 the Born approximation of perturbative QCD (see \fig{mpsi}).  
  \eq{HH1} simply  states that we can consider the QCD parton cascade of
 \eq{PC1} generated by the dipole of the size $r$ for 
the c.m.f.  rapidities from 0 to $\h Y$, and the same cascade for 
the dipole
 of the size $R$, for the rapidities from 0 to $ -\h Y  $.

 Generally speaking, for the dense system of partons at $Y \, =\,0$
 $n$-dipoles from upper cascade could  interact with $m$ dipoles 
from the low cascade, with the amplitude $N^m_n\Lb\{r_i\},\{r'_j\}\Rb$
\cite{LELU2}.
 In \eq{HH1} we assume that the system of dipoles that has been created
 at $Y =0$ is not very dense. In this case 
 \be \label{HH10}
 N^m_n\Lb\{r_i\},\{r'_j\}\Rb\,\,=\,\,\delta_{n,m}\prod^n_{j =1}\,\Lb - 1\Rb^{n - 1}N^{\rm BA}\Lb r_i, r'_i,b''_i\Rb
 \ee
  and after integration over $\{r_1\}$ and $\{r'_j\}$, the scattering
 amplitude can be reduced to a system of enhanced BFKL Pomeron
 diagrams, which are shown in \fig{mpsi}-b.
 
 The average number of dipoles at $Y = 0$ are determined by the 
inclusive cross section, which is given by the diagram of
 \fig{mpsi}-c and which can be written at $y \to 0$  as
 follows\cite{KTINC}:
 \be \label{HH2}
\frac{d \sigma}{d y \,d^2 p_{T}} = \frac{2C_F}{\alpha_s (2\pi)^4}\,\frac{1}{p^2_T} \int d^2 \vec r_T\,e^{i \vec{p}_T\cdot \vec{r}_T}\!\!\int\!\! d^2 b\,\nabla^2_T\,N^{\rm \mbox{\tiny  BFKL}}\Lb \h Y ;r,  r_T; b \Rb\,\int \!\!d^2 B \,\nabla^2_T\,N^{\rm\mbox{\tiny  BFKL}}\Lb y_2 = -\h Y ;R,  r_T; B \Rb
\ee
The average number of dipoles that enters the multiplicity
 distribution of \eq{NDIST} is equal $ \bar{n} \,=\,N\,=\int
 \frac{d^2 p_T}{(2 \pi)^2} \frac{d \sigma}{d y \,d^2 p_{T}}\Big{/}
 \sigma_{in} \,\propto\,\exp\Lb \Delta_{\rm BFKL}\,Y\Rb$\footnote{$
 \Delta_{\rm BFKL}$ is the intercept of the BFKL Pomeron.}
only if we assume that $\sigma_{in} \,\sim {\rm Const}$. 
Indeed, the enhanced diagrams of \fig{mpsi}-b lead to the
 inelastic cross section which is constant at high energy.

 \subsection{Hadron - hadron collisions}

\begin{figure}[ht]
     \begin{center}
     \begin{tabular}{c c c}
  \includegraphics[width= 0.4\textwidth]{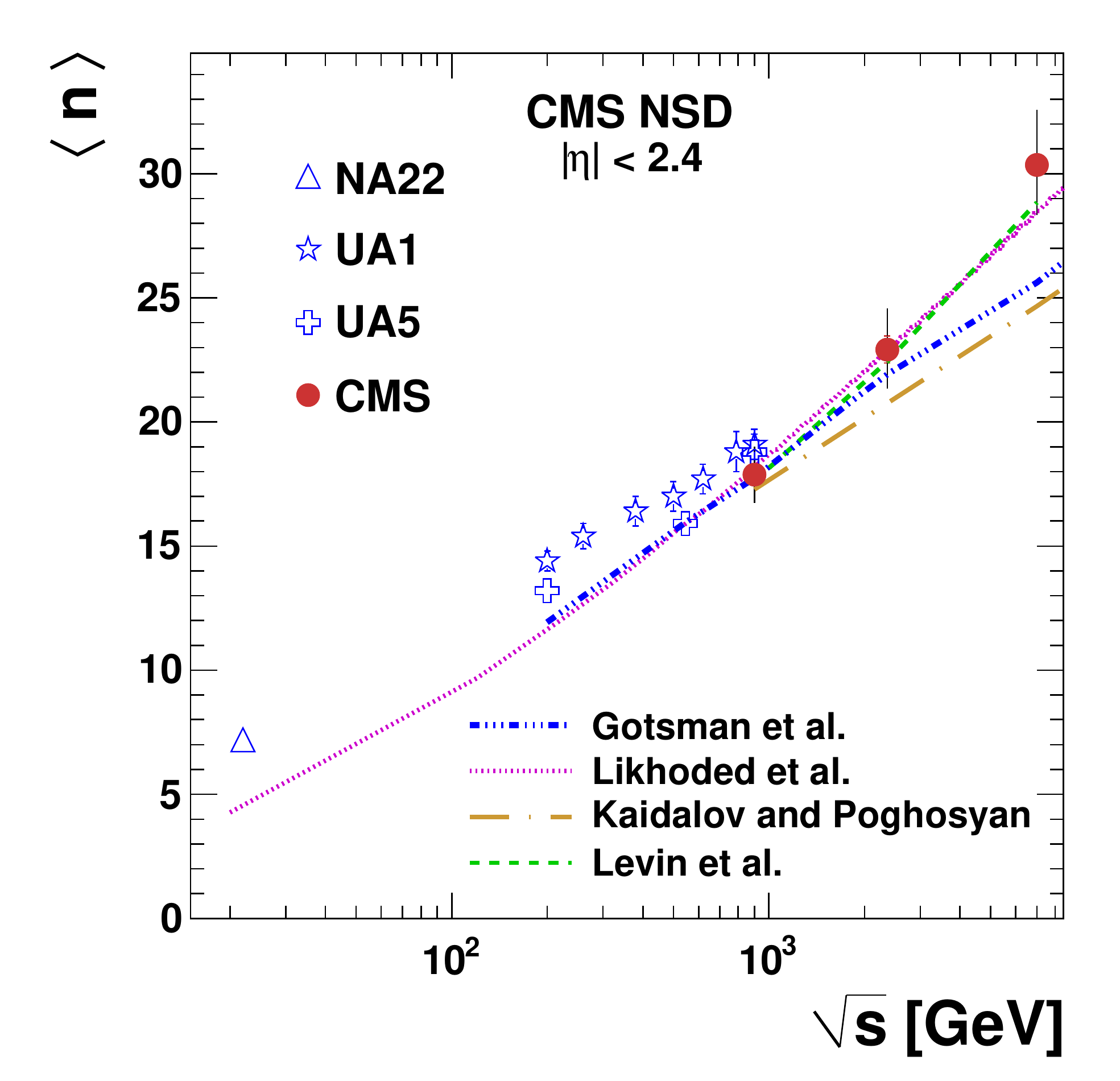}  &~~~~~~~~~~&   \includegraphics[width= 0.35\textwidth]{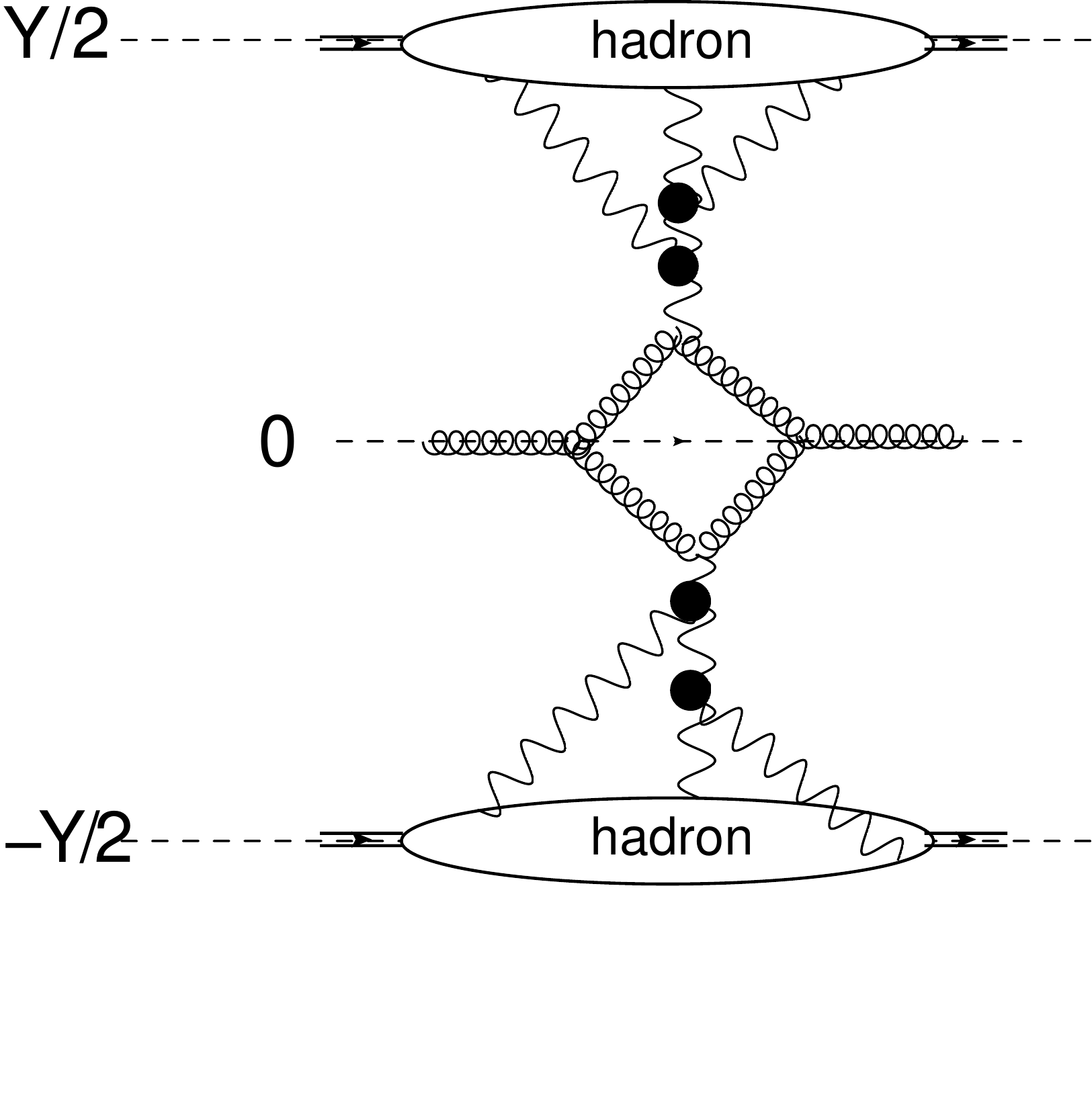}  \\
  \fig{dndyexp}-a & & \fig{dndyexp}-b\\
  \end{tabular}
   \end{center}    
      \caption{\fig{dndyexp}-a: The comparison of the average multiplicities
 in proton-proton collisions at $|\eta\,\leq \,2.4$\cite{CMSMULT} with the
 theoretical prediction\cite{GKLM,LLN,KAPO,LERE}. The figure is taken from
 Ref.\cite{CMSMULT}. The CGC prediction is marked by {\it Levin et al.} 
 and they are taken from Ref.\cite{LERE}.\fig{dndyexp}-b:
  The Mueller diagram \cite{MUDIA} for the inclusive
 production in CGC/saturation approach. The wavy lines are the BFKL
 Pomerons. The helical lines denote the gluons. The black blobs stand 
for the triple Pomeron vertices.}
\label{dndyexp}
   \end{figure}

   The first idea is to view a  hadron as a dilute system of dipoles and 
use
 \eq{HH2} for the average multiplicity, 
   together with the multiplicity  distribution of \eq{NDIST}. However, the
  energy dependence of the mean multiplicity from \eq{HH2}
 ($\bar{n} \,\propto\,\exp\Lb \Delta_{\rm BFKL}\,Y\Rb$ does
 not describe the experimental data (see \fig{dndyexp}). However,
  the experimental dependence of the mean multiplicity 
   on energy can be parameterized as $\bar{n} \,\propto\,\exp\Lb
 \lambda\,Y\Rb$,   but with the value of $\lambda \,=\,0.1 \div
 0.2$\cite{GKLM,LLN,KAPO}, which is far too small for $\Delta_{\rm BFKL}$.
 However, 
   this power is  close to the experimental behaviour of the deep inelastic
 structure function. Therefore, the main assumption of Ref.\cite{KHLE},  
 that $N \sim xG(x,Q^2)$ , does not contradict the experimental data at
 least on the qualitative level\cite{TKU}. 
   
On the other hand, the experimental data can be described in the
 framework of the CGC/saturation approach in which $N^{\rm 
\mbox{\tiny  BFKL}}$ were replaced by $N^{\rm \mbox{\tiny BK}}$\cite{LERE}.
Hence, we cannot view hadrons as the dilute system of 
 dipoles, but rather have to consider them as the
a dense system of dipoles.  For such a situation we expect
 that $\bar{n} \,\propto \,Q^2_s(Y)/\bas$ (see Refs.\cite{KOLEB,KLN,DKLN,
LERE,LAPPI}.  Therefore the entanglement  entropy in this case 
\be \label{HH3}
S_{\rm parton\,cascade}\,\,=\,\,\ln \bar{n} \,\,=\,\,\ln
 \Lb Q^2_s(Y)/Q^2_s\Lb Y=0\Rb\Rb
\ee

\begin{figure}[ht]
     \begin{center}
  \includegraphics[width= \textwidth]{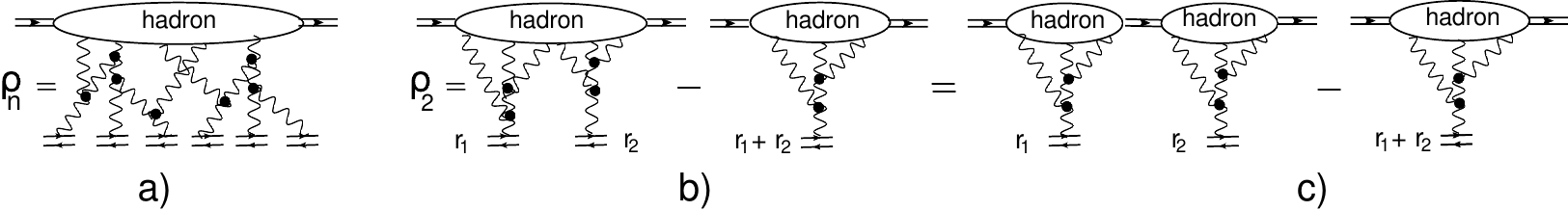}  
     \end{center}    
      \caption{\fig{rho}-a: The BFKL Pomeron diagrams for
 $\rho_n$. \fig{rho}-b and \fig{rho}-c: The Pomeron diagrams
 for $\rho_2$. The wavy lines denote the BFKL Pomerons, The
 black circles stand for triple Pomeron vertices. The arrows
  describe quarks and anti-quarks. }
\label{rho}
   \end{figure}

 Note that for CGC approach of section III, the
 average multiplicity turns out to be proportional $\mu^2\,\,
 \sim\,\, Q^2_s$, if the energy evolution is  taken into
 account (see Ref.\cite{KOLEB,GIJMV} for review). Frankly 
speaking, we do not have a theoretical tool to treat the 
dense-dense system scattering. For a general set of  
diagrams (see \fig{rho}-a) we cannot use neither the
 Hamiltotian of Ref.\cite{AKLL1}, nor other theoretical methods.
 Therefore, our suggestion to use the multiplicity distribution
 of \eq{NDIST} with $\bar{n}$ determined by \eq{HH2} with 
 $N^{\rm \mbox{\tiny BK}}$ is a {\it conjecture}. However,
 we can claim, that  $\rho_2\,\,=\,\,\rho_1(r_1)\rho_1(r_2)
 \,-\,\rho_1\Lb \vec{r}_1 + \vec{r}_2\Rb$ (see \fig{rho}-b
 and \fig{rho}-c) , on the same theoretical grounds as the
 derivation of BK equation\cite{B,K} since,  the fact that
  the amplitude for two BFKL Pomeron production is equal to 
$\Lb N^{\rm \mbox{\tiny BK}}\Rb^2$, is used in the derivation
 of the BK equations.

 \subsection{Comparison with the experimental data}
  In  order to compare the 
parton cascade with the
 experimental data, we first need  
  to establish a relation between the  multiplicity of
 hadrons with the multiplicities of partons, in the QCD parton cascade. 
Based on ``parton liberation" picture \cite{MULIB} and on the " local
 parton-hadron duality" \cite{LHPD} we assume that $S_{\rm parton
 \,cascade} \,\,=\,\,S_{hadrons}$. In other words, we  suggest
 that there is no 
   substantial entropy increase during the transformation of partons
 to hadrons.  This relation is our hypothesis about confinement of
 quarks and gluons,  and it has  support in the fact, that the 
value of the entropy corresponds to 
   a maximally entangled, equipartitioned state at a relatively
 modest  average multiplicity of  around  $N = 3 \div 6$.
 
 Hence, we use \eq{NDIST} for the hadron multiplicity distribution
 replacing $\bar{n} = N $ of partons, by $\bar{n} $ of hadrons.

Following Ref.\cite{KHLE}, we estimate, using \eq{NDIST}, the value of the 
 the cumulants 
\be\label{CUMDEF}
C_q\,\, =\frac{ <n^q>}{<n>^q} ,
\ee
 where $<\dots >$ denotes the average over the  distribution in hadron
 multiplicity $n$. These quantities can be readily computed using
 \eq{NDIST}(see Ref.\cite{KHLE}). The result of these estimates
 are the following:

\bea\label{CQ1}
C_2 &=& 2 - 1/\bar{n};~~~~ C_3\,\,=\,\,\frac{ 6 (\bar{n} - 1)\bar{n} +1}{\bar{n}^2};\nn\\
C_4&=& \frac{(12 \bar{n}(\bar{n} - 1) +1)(2 \bar{n} -1)}{\bar{n}^3};~~~~~ C_5\,\,=\,\, \frac{(\bar{n} - 1)( 120 \bar{n}^2 (\bar{n} - 1) + 30 \bar{n} )+1}{\bar{n}^4} .
\eea

Using the experimental multiplicity in the rapidity window
 $|\eta | \leq 0.5$ equal\cite{CMSMULT} to $\bar{n}
 = 6.33 \pm 0.46$ at W=7 \,TeV and $\bar{n} \,=\,3.72\,\pm\,0.23$ at
  W= 0.9\, TeV 
 we get from (\eq{CQ1}) the following predictions for the cumulants:
 $C_2 \simeq 1.83 (1.73)$, $C_3 \simeq\,5.08(4.46)$, $C_4 
\simeq 18.6(15.31)$ and $C_5 \simeq 85.7(65.75)$. We put in
  parentheses the values at $W = 0.9\,TeV$.  The CMS experiments
 give (see Fig.6-b of Ref.\cite{CMSMULT}):
 $C_2^{\rm exp} = 2.0 \pm 0.05$, 
$C_3^{\rm exp} = 5.9 \pm 0.6$, 
$C_4^{\rm exp} = 21 \pm 2$, and 
$C_5^{\rm exp} = 90 \pm 19$. Therefore, our estimates are in 
 reasonably good agreement with the data, indicating that the
 parton distributions are close to the hadronic ones. Taking the
 limit of $\bar{n}\,\to\,\infty$ we obtain the maximal values
 for the cumulants :
 $C_2 = 2$, $C_3= 6$, $C_4=24$ and $C_5 = 120$ as a prediction
 for  asymptotically high energies. Comparing these numbers
 to the experimental values  listed above, we see that the
 multiplicity distribution measured at $\sqrt{s} = 7$ TeV
 is already quite close to the expected asymptotic form.

\begin{figure}[ht]
     \begin{center}
     \begin{tabular}{c c c}
  \includegraphics[width= 0.45\textwidth]{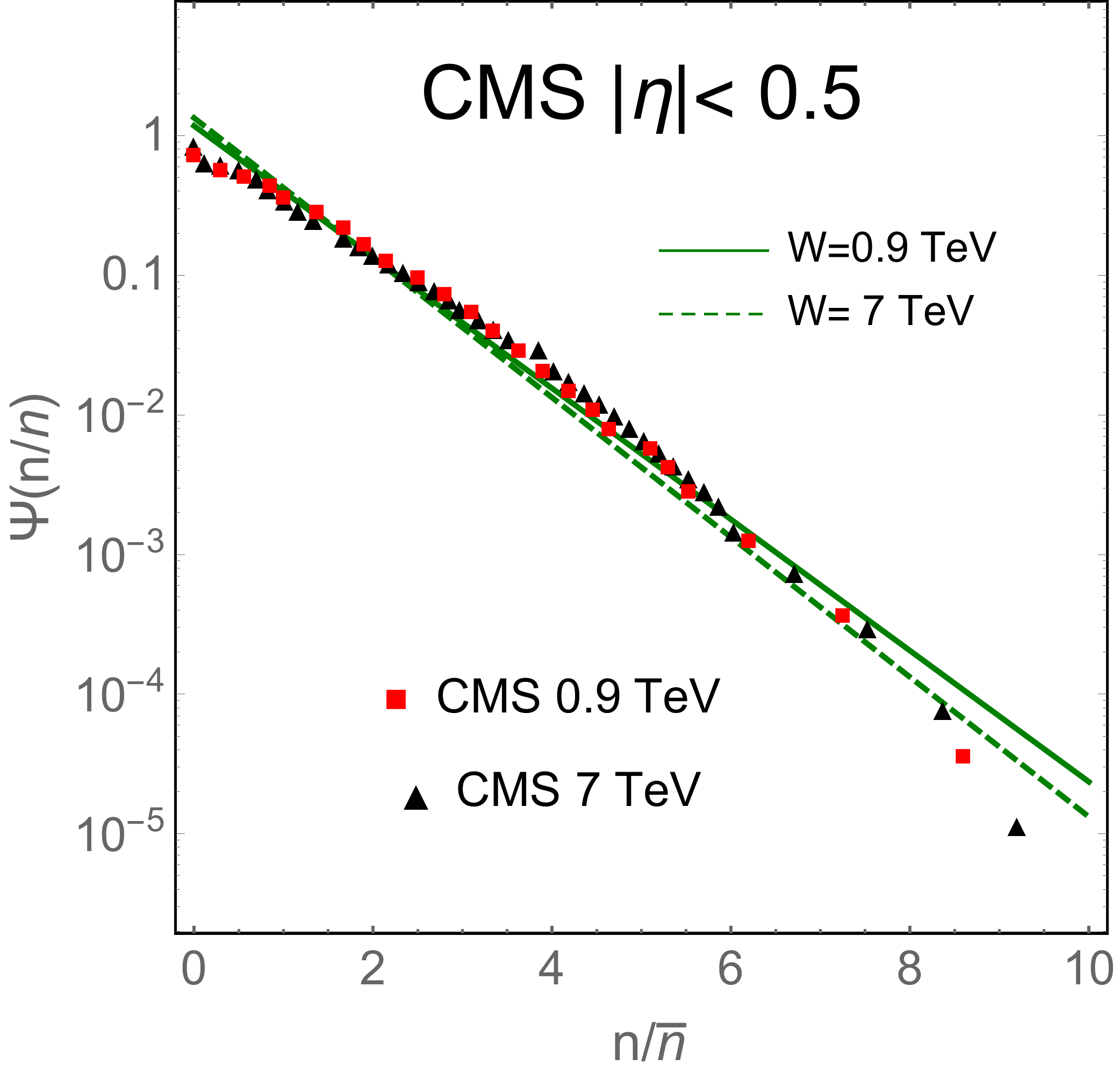}  &~~~~~~~~~~&   \includegraphics[width= 0.45\textwidth]{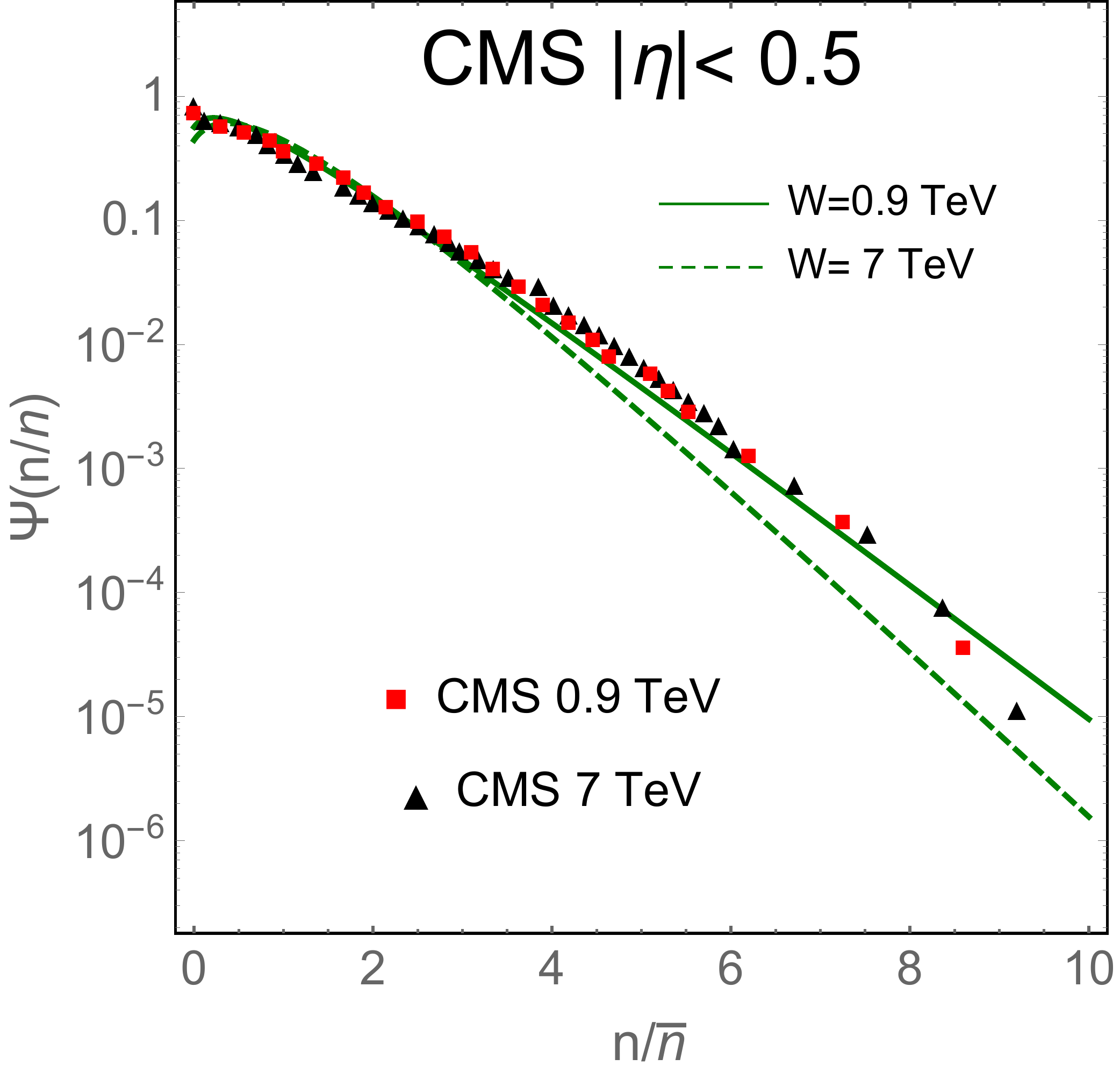}  \\
  \fig{kno1}-a & & \fig{kno1}-b\\
  \end{tabular}
   \end{center}    
      \caption{KNO function $\Lb \frac{\sigma_n}{\sigma_{\rm in}}\,\,=\,\,\frac{1}{\bar{n}} \Psi\Lb \frac{n}{\bar{n}}\Rb\Rb$
      versus $z = n/\bar{n}$ for rapidity window
 $| \eta|\,<\,0.5$. \fig{kno1}-a: Comparison with
 prediction of \eq{NDIST}.\fig{kno1}-b: Comparison
 with the  negative binomial distribution of \eq{NBD},
 in which we estimated parameter $r$ using $\rho_2\,\,
=\,\,\rho_1(r_1)\rho_1(r_2) \,-\,\rho_1\Lb \vec{r}_1 + \vec{r}_2\Rb$ . }
\label{kno1}
   \end{figure}

\begin{figure}[ht]
     \begin{center}
     \begin{tabular}{c c c}
  \includegraphics[width= 0.45\textwidth]{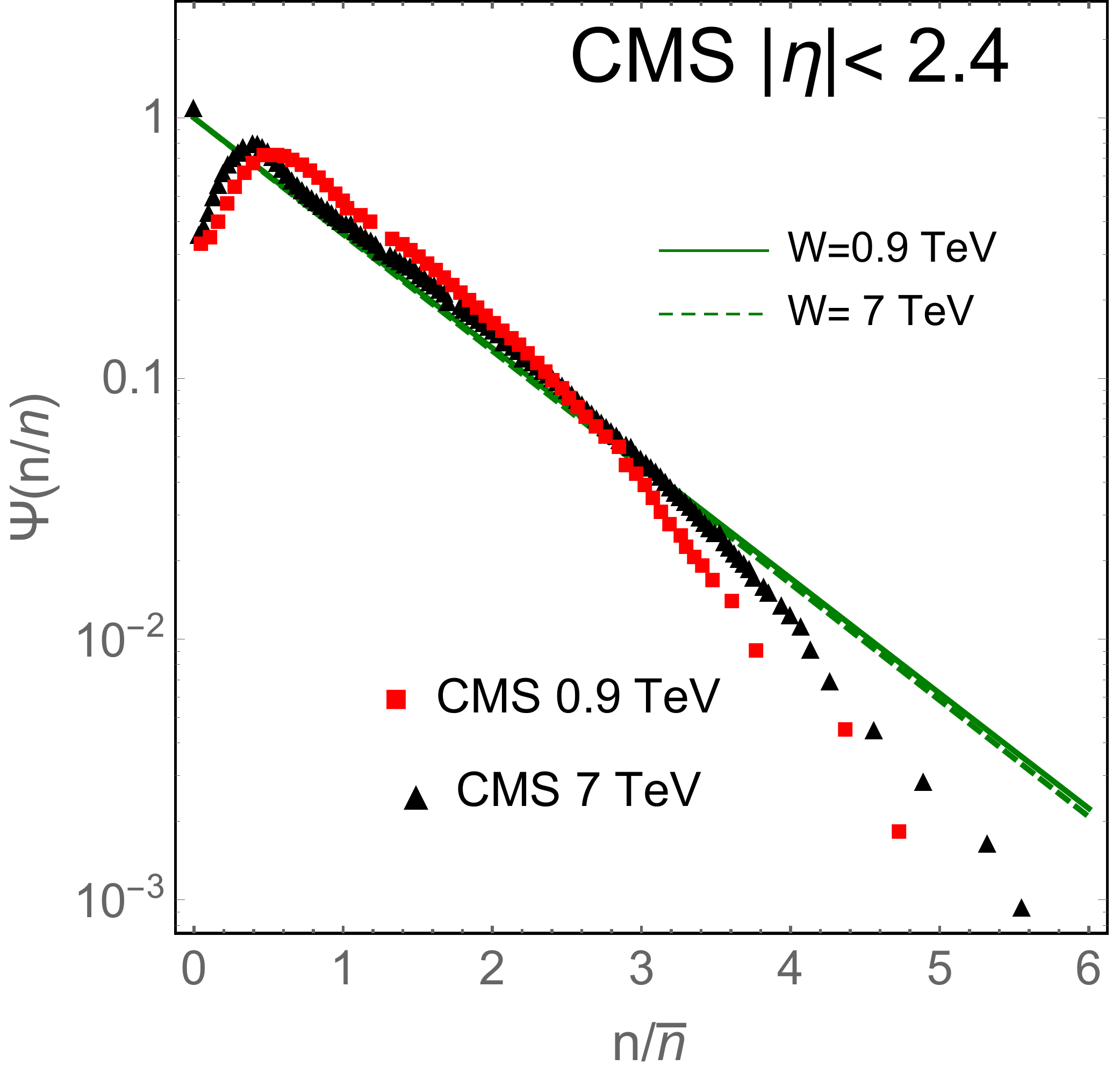}  &~~~~~~~~~~&   \includegraphics[width= 0.45\textwidth]{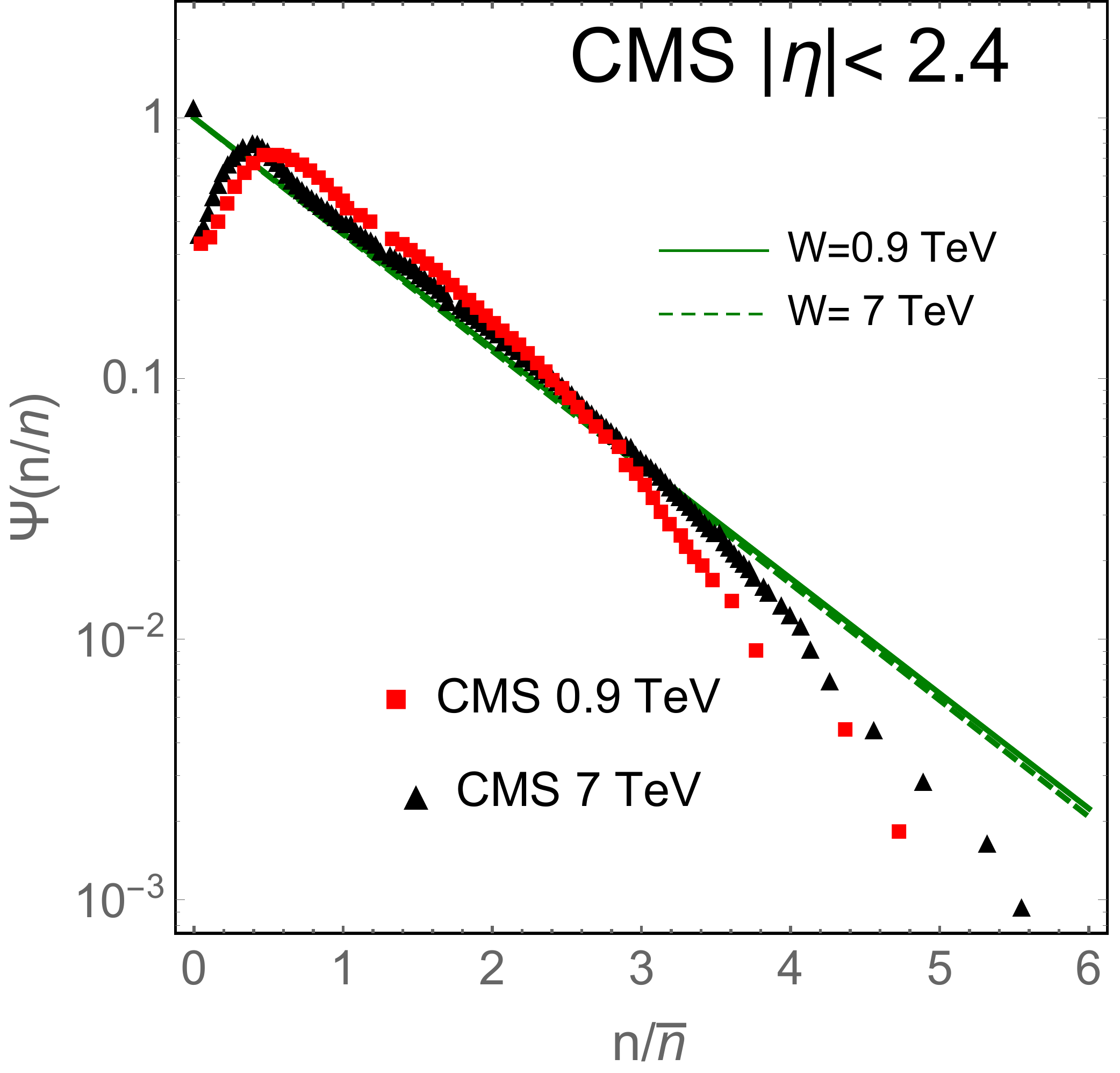}  \\
  \fig{kno2}-a & & \fig{kno2}-b\\
  \end{tabular}
   \end{center}    
      \caption{KNO function $\Lb \frac{\sigma_n}{\sigma_{\rm in}}\,\,
=\,\,\frac{1}{\bar{n}} \Psi\Lb \frac{n}{\bar{n}}\Rb\Rb$
      versus $z = n/\bar{n}$ for rapidity window $| \eta| \,<\,2.4$. 
\fig{kno2}-a: Comparison with prediction of \eq{NDIST}.\fig{kno2}-b: 
Comparison with the  negative binomial distribution of \eq{NBD}, in
 which we estimated parameter $r$ using $\rho_2\,\,=\,\,\rho_1(r_1)\rho_1(r_2) \,-\,\rho_1\Lb \vec{r}_1 + \vec{r}_2\Rb$ . }
\label{kno2}
   \end{figure}

 In \fig{kno1} and \fig{kno2} we plot the multiplicity dependence
 in the form of the KNO scaling function\cite{KNO}: 
 \be \label{HH4}
  \frac{\sigma_n}{\sigma_{\rm in}}\,\,=\,\,\frac{1}{\bar{n}}\Psi\Lb z\,\,=\,\, \frac{n}{\bar{n}}\Rb
  \ee
  One can see that at large $\bar{n}$ the distribution of \eq{NDIST}
 leads to
  \be\label{HH5}
   \Psi\Lb z =  \frac{n}{\bar{n}}\Rb\,\,\xrightarrow{\bar{n} 
\,\gg\,1}\,\,\,e^{- z}\Lb 1\,+\,\frac{1}{\bar{n}} \,-\,\frac{z}{2\,\bar{n}}\Rb
   \ee
    and shows  KNO scaling for $\bar{n}\,\gg\,z$. As far as
 we know, this is the first time that  KNO scaling appears at
 ultra high energies on theoretical grounds for hadron-hadron
 collisions. At least, in the framework of the Pomeron 
calculus\cite{GRIBPC}  KNO scaling is expected only 
 for the intermediate range of energy\cite{ABKA,MAWA}.

 In \fig{kno1}-a and \fig{kno2}-a we  compare  the CMS
 data\cite{CMSMULT} with the multiplicity distribution of
 \eq{NDIST}. In spite of the good agreement at small $z$, one 
can see two major qualitative disagreements:(i)  KNO 
scaling works better for $|\eta| \,<\,0.5$ than at 
$|\eta |<2.4$ in the 
data, while \eq{NDIST} predicts a different behaviour;
 and (ii) \eq{NDIST} predicts larger cross section in
 the region of large $z$, than  is observed experimentally.
  The first disagreement is intimately related to the small
 value of $\bar{n}$ at $|\eta| \,<\,0.5$. In the framework 
of our approach  the violation of KNO scaling will  not be 
 seen if we take $N\,=\,5.8$\cite{PDG}  at $\eta=0$, instead
 of $6.33 = 30.4/4.8$,  which we used in these figures. The
 second disagreement is of a principle nature. As we have discussed
 we use \eq{HH10},which is based on the assumption, that we do not have
 a dense system of parton at $Y=0$.
 Certainly, such an assumption is not correct for  events with
 large multiplicities.  Therefore, this disagreement can be considered
 as an additional argument in the attempts to build a generalization of
 CGC approach to describe the hadron-hadron collisions. 
 
 In \fig{kno1}-b and \fig{kno2}-b we use the negative 
 binomial distribution of \eq{NBD} in which we fixed 
the parameter $r$ from the moment $\Big{<} \frac{n(n-1)}{2}\Big{>}$
 , which is estimated using $\rho_2\,\,=\,\,\rho_1(r_1)\rho_1(r_2)
 \,-\,\rho_1\Lb \vec{r}_1 + \vec{r}_2\Rb$ (see \fig{rho}). One can
 see that this distribution has the same characteristic features as \eq{NDIST}.

 \subsection{Back to QCD motivated parton model}
 In the previous section we  inferred, that comparison with the
 experimental data  indicates, that we cannot use \eq{HH10}, which
 stems from the assumption, that the system of partons, which is
 produced at $Y=0$,  is rather dilute. Unfortunately,  we have not
   developed a theoretical approach in the framework of QCD to treat
 this problem. However, in the QCD parton model, that we have
 discussed above, a breakthrough has been made by   \cite{KLL},
 and  an  approach has been  constructed,  that sums all  pomeron 
diagrams
 of the most general type (see \fig{mpsi} and \fig{rho} for examples).
 
  In Ref.\cite{KLL} a new Hamiltonian is suggested , which has the form
 
 \be\label{HPM}
{\cal H}_{\rm PM}=-\frac{1}{\gamma}\bar{ P}\,P\ee
where PM stands for `` parton model''. $\bar P$ and $P$ are the BFKL
 Pomeron field in the model, where the sizes of the dipoles are fixed.
 This Hamiltonian in the limit of small $\bar P$   reproduces the
 BK Hamiltonian 
( see Ref.\cite{KLL}  and below for details). This condition is the most
 important one for fixing the form of
${\cal H}_{\rm PM}$.  The second of such conditions is that this
 Hamiltonian satisfies both $t$ and $s$ channel unitarity. 
 $\gamma $ in \eq{HPM}  has the physical meaning of  the 
dipole-dipole scattering amplitude in the Born approximation
 of perturbative QCD and, being such,  it is naturally small and of   
 order $\bas$.  

The most important ingredient of this approach is the generalization
 of the commutation relation, which has the form:
\be\label{CRCOR}
\Big(1\,\,-\,\,P\Big)\Big(1\,\,-\,\,\bar P \Big)\,\,=\,\,(1-\gamma)\Big(1\,\,-\,\,\bar P\Big) \Big(1\,\,-\,\,P\Big)
\ee
\eq{CRCOR}  gives the correct factor $(1-\gamma)^{\bar n}$ that includes
 all multiple scattering corrections, while all the dipoles remain intact,
 and can subsequently scatter on  an  additional projectile or on target 
dipoles.
For small $\gamma$,  and in the regime where  $P$ and $\bar P$ are
 small themselves, we obtain
\be
[P,\bar P]=-\gamma +...
\ee
consistent with our original expression. One can see that these 
commutation relations take into account the interaction of one
 dipole with many other partons and, therefore, we are going
 beyond  the approximation, which is given by \eq{HH10}.
 Concluding this brief  outline of this approach, we see, 
that for the first time we have a simple theory in which we can 
describe the interactions of dilute-dilute parton system scattering,
 as well as dilute-dense and dense-dense system  interactions.

For ${\cal H}_{\rm PM}$ the cascade equation takes the form (see
 Eq. 5.8 of Ref.\cite{KLL}):
\be \label{EQNPM}
 \frac{d P_n\Lb Y\Rb}{d Y}\,\,=\,\, - \frac{\Delta}{\gamma}\Lb 1 - \Big( 1\,-\,\gamma\Rb^n\Big)   P_n\Lb Y\Rb \,\,+\,\,\frac{\Delta}{\gamma}\Big( 1 - \Lb 1\,-\,\gamma\Rb^{n - 1}\Big) P_{n-1}\Lb Y .\Rb
\ee
For small $n$ ($ \gamma\,n\,<\,1$) one can see, that \eq{EQNPM} 
reduces to \eq{EQPM}. Hence, for such small $n$ we have the multiplicity
 distribution  of \eq{PNPM} with $< n> = e^{\Delta\,Y}$.
However, at large $n$ \eq{EQNPM}  has the form 
\be \label{EQNPM1}
 \frac{d P_n\Lb Y\Rb}{d Y}\,\,=\,\, - \frac{\Delta}{\gamma}   P_n\Lb Y\Rb \,\,+\,\,\frac{\Delta}{\gamma} P_{n-1}\Lb Y \Rb
\ee
We will show below that this equation gives the Poisson distribution
 with $< n> \,=\,\frac{\Delta}{\gamma}Y$. Therefore, as we have guessed
 the interaction of one dipole with many dipoles at $Y=0$ in \fig{mpsi}
 would lead to far fewer  multiplicities than \eq{PNPM}. Using 
Laplace
 transform of \eq{LT} and following the pattern, described in
 section II-C, we obtain the solution in $\omega$-representation:

\be \label{SOLNPM}
 P_n\Lb \omega\Rb\,\,=\,\,\,\frac{1}{\omega_1} \prod^n_{m = 1}\frac{\omega_m}{\omega\,\,+\,\,\omega_m}
 \ee
 where $\omega_m\,\,=\,\,\frac{\Delta}{\gamma}\Big( 1 - \Lb 1\,-\,
\gamma\Rb^{m}\Big)$. 
 
 We have not found an  elegant form for the inverse Laplace transform,
 but  we can 	 see   the main qualitative features of this 
solution,  assuming that for $n\,<\,n_0$ with $\gamma \,n_0 \approx\,1$
 we have $\omega_m\,=\,m\,\Delta$, but for $n \,>\,n_0$ $\omega_m\,=
\,\frac{\Delta}{\gamma}$. In this approach we obtain:
 \begin{subequations}
   \bea 
 n\,<\,n_0  &  P_n\Lb Y\Rb \,\,=\,\,&e^{- \Delta\,Y} \Big( 1 \,\,-\,\,e^{ - \Delta\,Y}\Big)^{n - 1} ;\label{SOLNPM1}\\
 n\,>\,n_0 & P_n\Lb Y\Rb \,\,=\,\,& \int^Y d Y'  e^{- \Delta\,\Lb Y - Y'\Rb} \Big( 1 \,\,-\,\,e^{ - \Delta\,\Lb Y - Y'\Rb}\Big)^{n_0 -1} \,\underbrace{ \,e^{ - \frac{\Delta}{\gamma} Y'} \frac{\Lb \frac{\Delta}{\gamma} Y'\Rb^{n - n_0}}{\Lb n - n_0\Rb!}}_{\rm Poisson\, distribution};\label{SOLNPM2}
 \eea
\end{subequations}  
The Poisson distribution in \eq{SOLNPM2} is the inverse Laplace transform
 of 
\be \label{NPM1}
\frac{\Lb  \frac{\Delta}{\gamma} \Rb^{n - n_0}}{\Lb \omega\,\,+\,\,\frac{\Delta}{\gamma}\Rb^{n - n_0 + 1}}
\ee

     \begin{figure}[ht]
     \begin{center}
     \begin{tabular}{c c c}
  \includegraphics[width= 0.3\textwidth]{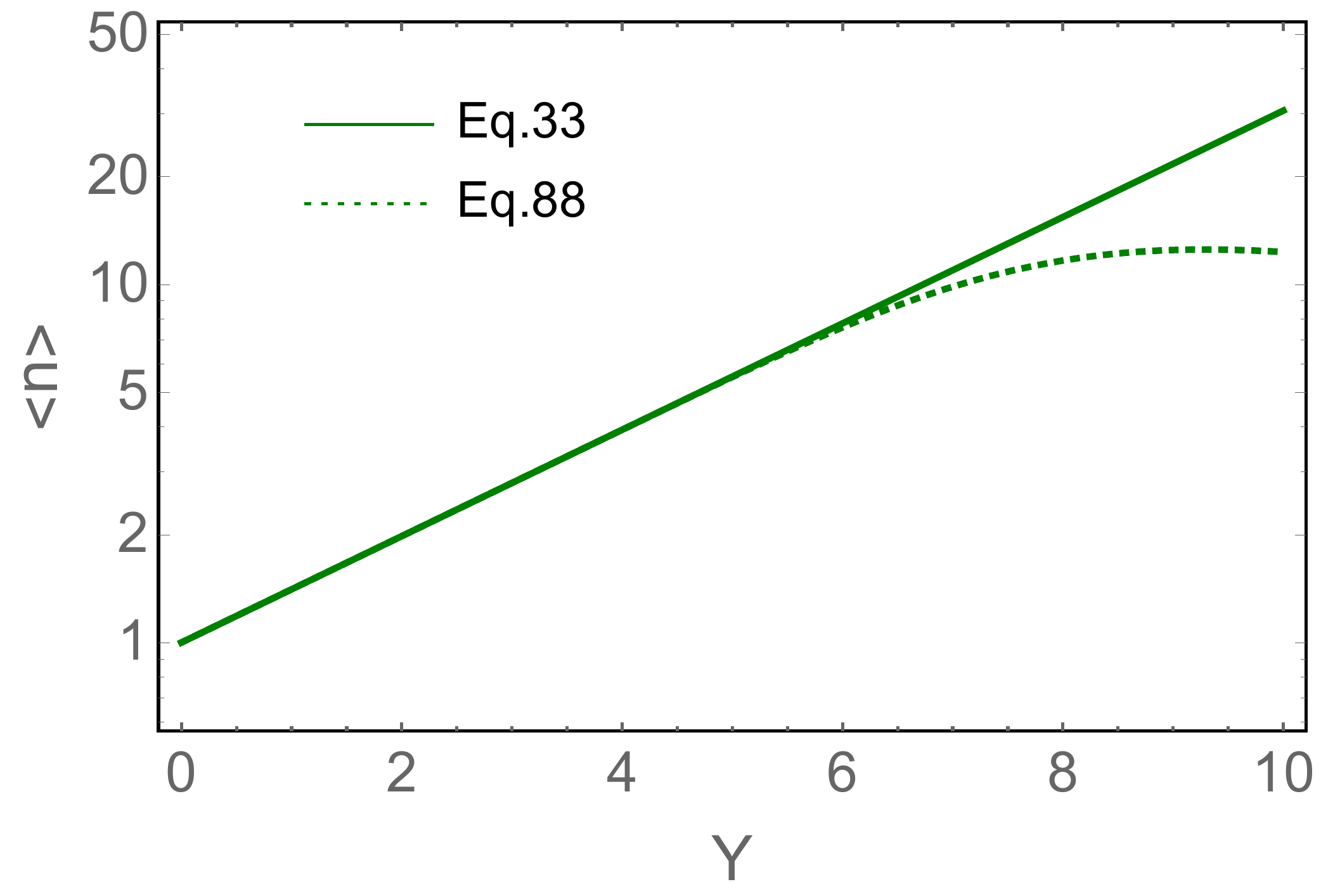}  &   \includegraphics[width= 0.315\textwidth]{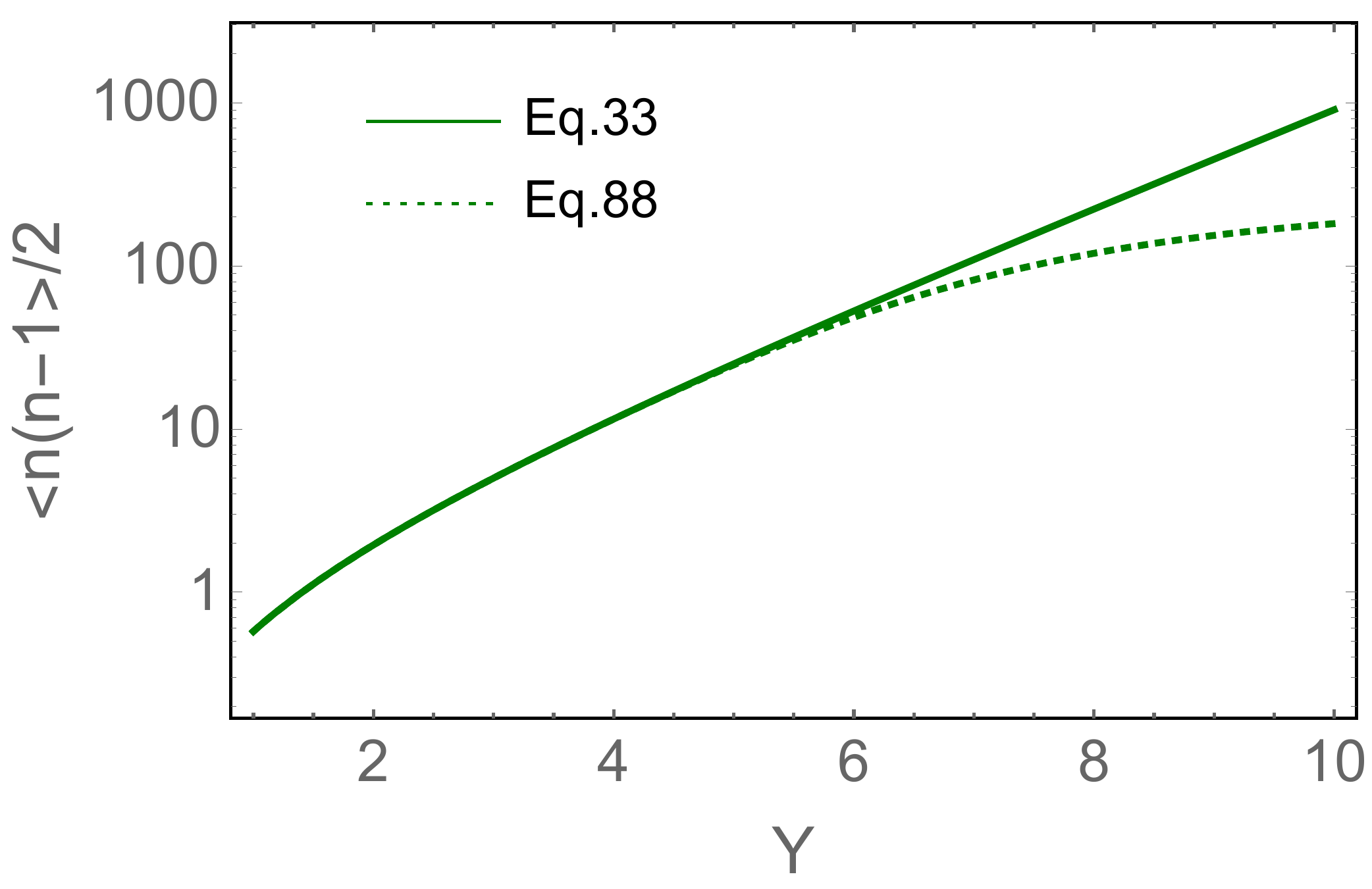} &  \includegraphics[width= 0.315\textwidth]{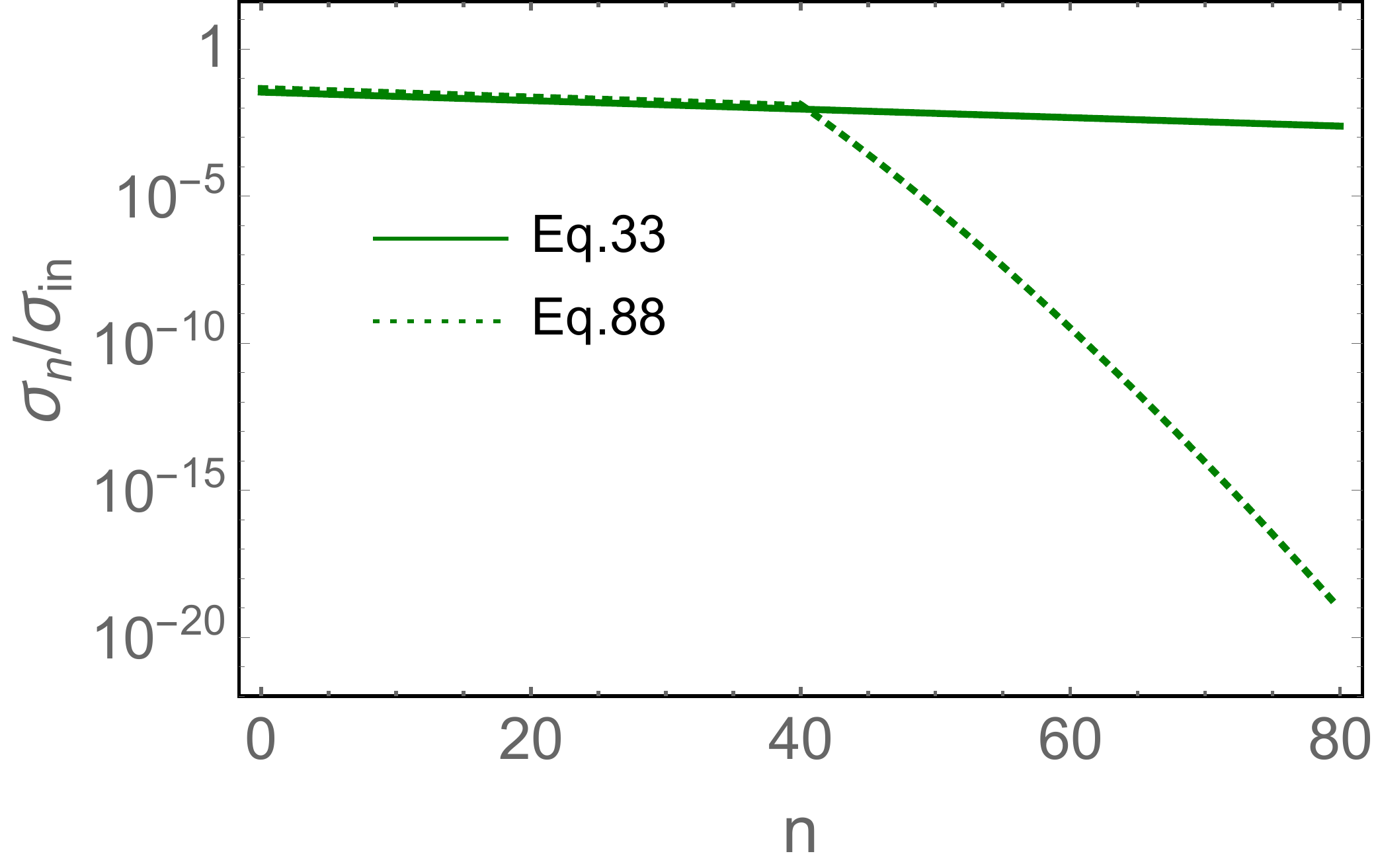} \\
  \fig{moddist}-a &\fig{moddist}-b&\fig{moddist}-c\\
  \end{tabular}
   \end{center}    
      \caption{ Comparison of the multiplicity distribution, given
 by \eq{NDIST}  with the modified distribution of \eq{SOLNPM1} and
 \eq{SOLNPM2}.}
\label{moddist}
   \end{figure}

In \fig{moddist} we compare this multiplicity distribution with
 \eq{NDIST}. One can see that at large multiplicities the modified
 distribution of \eq{SOLNPM1} and \eq{SOLNPM2} lead to the suppression
 of the parton emission, as we expected. Of course, this
 modified distribution is very approximate, and can only be used
  to clarify the qualitative features  of the interaction
 of the partons in the exact approach.

\begin{figure}[ht]
     \begin{center}
\includegraphics[width= 0.45\textwidth]{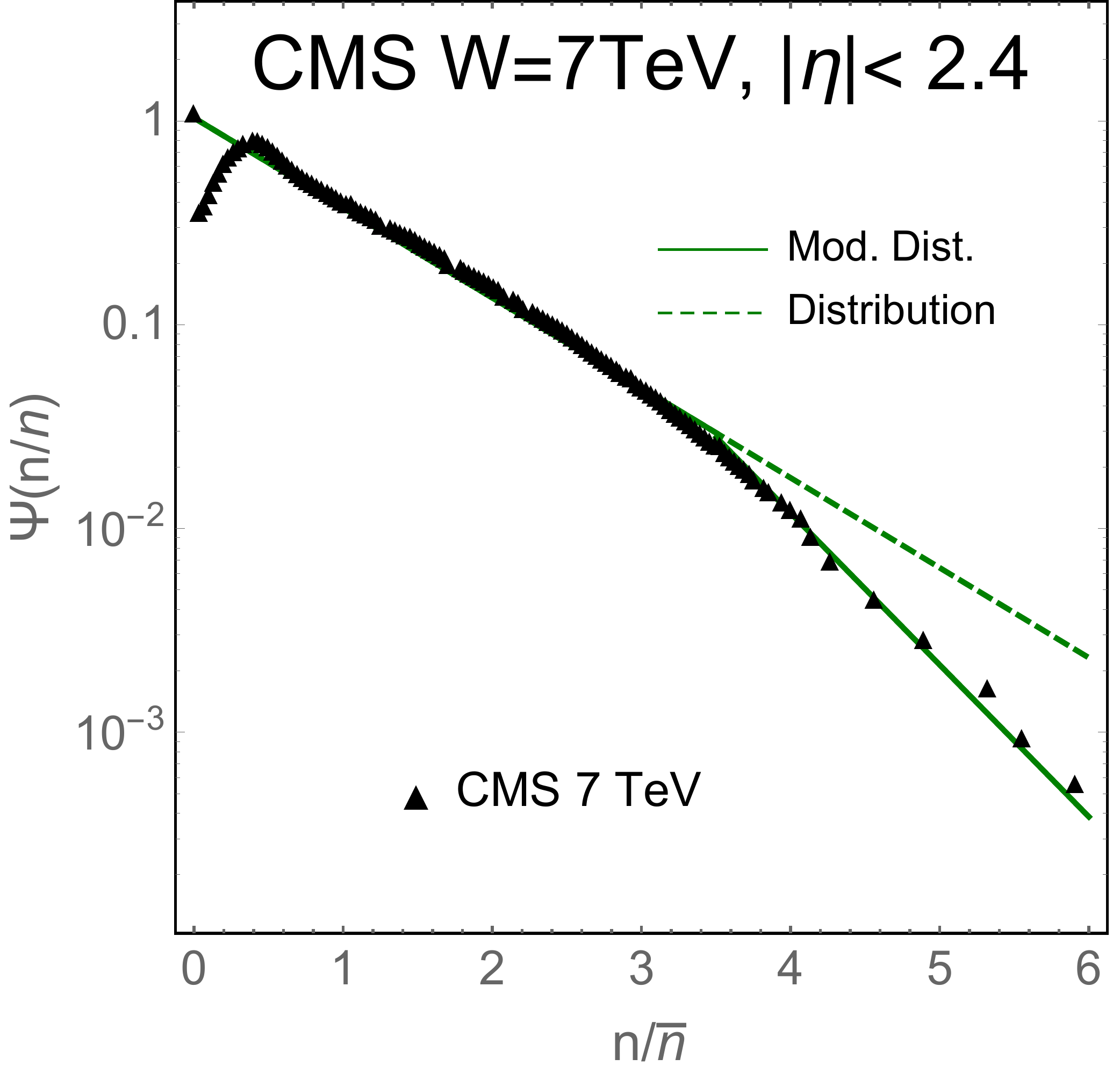}   
 \end{center}    
      \caption{KNO function $\Lb \frac{\sigma_n}{\sigma_{\rm in}}\,\,
=\,\,\frac{1}{\bar{n}} \Psi\Lb \frac{n}{\bar{n}}\Rb\Rb$
      versus $z = n/\bar{n}$ for rapidity window $| \eta| \,<\,2.4$. 
 Comparison between multiplicity distribution of \eq{NDIST} and modified
 distribution of \eq{NEWPN}, which was used for $z > 3.5$., $\gamma$ is
 equal to 0.15. The data is taken from Ref.\cite{CMSMULT}.  }
\label{knoexp}
   \end{figure}

To illustrate that
the parton cascade of \eq{EQNPM} is able to describe the experimental
 data. We calculate the first two $P_1$ and $P_2$, taking integral 
over $\omega$ in \eq{LT}:
\be \label{P12}
P_1\Lb Y \Rb\,\,=\,\,e^{\Delta\,Y}; ~~~~~~~~P_2\Lb Y\Rb\,\,=\,\,\frac{\omega_2}{\omega_2 - \omega_1}
e^{- \omega_1\,Y}\Big( 1\,\,-\,\,e^{ - \Lb \omega_2 - \omega_1\Rb Y}\Big)
\ee
where $\omega_2 \,-\,\omega_1\,=\,\Delta \,-\,\gamma \,\Delta\,\,<\,\,\Delta$. $P_2$ in our notation with $N\,\,=\,\,e^{\Delta\,Y} $ can be re-written in the form:
\be \label{NEWP2}
P_2\,\,=\,\,\frac{1}{N}\Lb 1\,\,-\,\,\Lb \frac{1}{N}\Rb^{1 - \gamma}\Rb
\ee
 Assuming that the multiplicity distribution has the form:
\be \label{NEWPN}
P_n\,\,=\,\,\frac{1}{N}\Lb 1\,\,-\,\,\Lb \frac{1}{N}\Rb^{1 - \gamma}\Rb^{n - 1}
\ee
 We can use this approximation, except   for very large $n$.
 In \fig{knoexp} we compare \eq{NEWPN} with the data. One can see that
 it   provides  quite a good description of the data. 

\section{Conclussions}
 
 As has been discussed in the introduction, this paper has two main 
results. 
  First, we prove  that in  QCD at high energies the multiplicity
 distribution has the form of \eq{NDIST}, which was discussed in
 Ref.\cite{KHLE}, in the framework of the parton model. We also show
 that the average number of gluons is not always related to  the gluon
 structure function, and can depend both on energy and the size of the
 dipoles. However, the entanglement entropy is equal to $S_{\rm partons}
 \,=\,\ln N$, where $N$ is the average number of partons, confirming that
 {\it  the partonic state at high  energy  is maximally entangled}
 \cite{KHLE}. In the case of DIS we prove that the average number
 of partons  is related to the gluon structure function,  only
 if we use  the BFKL evolution equation for  this structure
 function,  can we prove that  the multiplicity distribution
 has the form of \eq{NDIST}.
 
 Second, we developed an approach for hadron-hadron collisions
 in which we show that \eq{NDIST} correctly describes the
 Mueller-Patel-Salam-Iancy approach\cite{MUPA,MPSI}. We argued
 that actually at high energies the mean multiplicity is
 proportional to $Q^2_s\Lb Y\Rb$, leading to the entanglement entropy
 $\propto\,\ln Q^2_s(Y)$.
 
 We compared the multiplicity distribution with the experimental data.
 We described quite well the data at multiplicities $n \,<\,(3\div 4) < n>$,
 using \eq{NDIST},  but predict higher $\sigma_n$ for large multiplicities
 than has been seen experimentally. We conclude that this indicates
 that the assumptions ,  that at high energies we create a dilute
 system of partons in the c.m. rapidity $Y=0$,  and we can use \eq{HH10},
 is not correct, and we have to deal with a rather  dense system of 
partons.
 At the moment, we have not developed  theoretical tools to treat such
 a system. However, in  Ref.\cite{KLL}, an approach was developed for
 the parton model, which allow us to  theoretically treat such dense
 system of partons. We show that in this approach the production  of a
  system of partons with large multiplicities is suppressed in
 comparison with \eq{NDIST}, and we are able to describe the  
experimental data.

\section{Acknowledgements}
   We thank our colleagues at Tel Aviv university and UTFSM for
 encouraging discussions.  This research was supported  by 
 ANID PIA/APOYO AFB180002 (Chile) and  Fondecyt (Chile) grants  
 1180118.


\begin{thebibliography}{99} \frenchspacing
  
  
\bibitem{KUT}
K.~Kutak,
{\it ``Gluon saturation and entropy production in proton?proton collisions,''}
Phys. Lett. B \textbf{705} (2011), 217-221
doi:10.1016/j.physletb.2011.09.113
[arXiv:1103.3654 [hep-ph]].


\bibitem{PES}
R.~Peschanski,
{\it ``Dynamical entropy of dense QCD states,''}
Phys. Rev. D \textbf{87} (2013) no.3, 034042
doi:10.1103/PhysRevD.87.034042
[arXiv:1211.6911 [hep-ph]].



\bibitem{KOLU1}
A.~Kovner and M.~Lublinsky,
{\it ``Entanglement entropy and entropy production in the Color Glass Condensate framework,''}
Phys. Rev. D \textbf{92} (2015) no.3, 034016
doi:10.1103/PhysRevD.92.034016
[arXiv:1506.05394 [hep-ph]].

\bibitem{PESE}
R.~Peschanski and S.~Seki,
{\it ``Entanglement Entropy of Scattering Particles,''}
Phys. Lett. B \textbf{758} (2016), 89-92
doi:10.1016/j.physletb.2016.04.063
[arXiv:1602.00720 [hep-th]].

\bibitem{KHLE}
D.~E.~Kharzeev and E.~M.~Levin,
{\it ``Deep inelastic scattering as a probe of entanglement,''}
Phys. Rev. D \textbf{95} (2017) no.11, 114008
doi:10.1103/PhysRevD.95.114008
[arXiv:1702.03489 [hep-ph]].

\bibitem{BAKH}
O.~Baker and D.~Kharzeev,
{\it ``Thermal radiation and entanglement in proton-proton collisions at energies available at the CERN Large Hadron Collider,''}
Phys. Rev. D \textbf{98} (2018) no.5, 054007
doi:10.1103/PhysRevD.98.054007
[arXiv:1712.04558 [hep-ph]].



\bibitem{BFV}
J.~Berges, S.~Floerchinger and R.~Venugopalan,
{\it ``Dynamics of entanglement in expanding quantum fields,''}
JHEP \textbf{04} (2018), 145
doi:10.1007/JHEP04(2018)145
[arXiv:1712.09362 [hep-th]].


\bibitem{HHXY}
Y.~Hagiwara, Y.~Hatta, B.~W.~Xiao and F.~Yuan,
{\it ``Classical and quantum entropy of parton distributions,''}
Phys. Rev. D \textbf{97} (2018) no.9, 094029
doi:10.1103/PhysRevD.97.094029
[arXiv:1801.00087 [hep-ph]].

\bibitem{KOV1}
N.~Armesto, F.~Dominguez, A.~Kovner, M.~Lublinsky and V.~Skokov,
{\it ``The Color Glass Condensate density matrix: Lindblad evolution, entanglement entropy and Wigner functional,''}
JHEP \textbf{05} (2019), 025
doi:10.1007/JHEP05(2019)025
[arXiv:1901.08080 [hep-ph]].

\bibitem{GOLE1}
E.~Gotsman and E.~Levin,
{\it ``Thermal radiation and inclusive production in the CGC/saturation approach at high energies,''}
Eur. Phys. J. C \textbf{79} (2019) no.5, 415
doi:10.1140/epjc/s10052-019-6923-0
[arXiv:1902.07923 [hep-ph]].


\bibitem{GOLE2}
E.~Gotsman and E.~Levin,
{\it ``Thermal radiation and inclusive production in the Kharzeev-Levin-Nardi model for ion-ion collisions,''}
Phys. Rev. D \textbf{100} (2019) no.3, 034013
doi:10.1103/PhysRevD.100.034013
[arXiv:1905.05167 [hep-ph]].
\bibitem{KOV2}
A.~Kovner, M.~Lublinsky and M.~Serino,
{\it ``Entanglement entropy, entropy production and time evolution in high energy QCD,''}
Phys. Lett. B \textbf{792} (2019), 4-15
doi:10.1016/j.physletb.2018.10.043
[arXiv:1806.01089 [hep-ph]].
\bibitem{NEWA}
D.~Neill and W.~J.~Waalewijn,
{\it ``Entropy of a Jet,''}
Phys. Rev. Lett. \textbf{123} (2019) no.14, 142001
doi:10.1103/PhysRevLett.123.142001
[arXiv:1811.01021 [hep-ph]].

\bibitem{LIZA}
Y.~Liu and I.~Zahed,
{\it ``Entanglement in Regge scattering using the AdS/CFT correspondence,''}
Phys. Rev. D \textbf{100} (2019) no.4, 046005
doi:10.1103/PhysRevD.100.046005
[arXiv:1803.09157 [hep-ph]].



\bibitem{FPV}
X.~Feal, C.~Pajares and R.~Vazquez,
{\it ``Thermal behavior and entanglement in Pb-Pb and p-p collisions,''}
Phys. Rev. C \textbf{99} (2019) no.1, 015205
doi:10.1103/PhysRevC.99.015205
[arXiv:1805.12444 [hep-ph]].

\bibitem{TKU}
Z.~Tu, D.~E.~Kharzeev and T.~Ullrich,
{\it ``Einstein-Podolsky-Rosen Paradox and Quantum Entanglement at Subnucleonic Scales,''}
Phys. Rev. Lett. \textbf{124} (2020) no.6, 062001
doi:10.1103/PhysRevLett.124.062001
[arXiv:1904.11974 [hep-ph]].

\bibitem{KOV3}
H.~Duan, C.~Akkaya, A.~Kovner and V.~V.~Skokov,{\it ``Entanglement, partial set of measurements, and diagonality of the density matrix in the parton model,''}
Phys. Rev. D \textbf{101} (2020) no.3, 036017
doi:10.1103/PhysRevD.101.036017
[arXiv:2001.01726 [hep-ph]].



\bibitem{BJ} 
  J.~D.~Bjorken,
  {\it ``Asymptotic Sum Rules at Infinite Momentum,''}
  Phys.\ Rev.\  {\bf 179}, 1547 (1969).
  doi:10.1103/PhysRev.179.1547
 

  \bibitem{FEYN}
 R.P.  Feynman, {\it  ``Very high-energy collisions of hadrons,''}
  Phys.\ Rev.\ Lett.\  {\bf 23}, 1415 (1969).
\,\,\,
  {\it ``Photon-hadron interactions,''}
  Reading 1972. {\it Photon-Hadron Interactions}, Reading, 1972.  
 
  \bibitem{BJP}
 J.D.  Bjorken and  E.A. Paschos,{\it``Inelastic Electron-Proton and 
$\gamma$ -Proton Scattering and the Structure of the Nucleon,"}
Phys. Rev. 185, 1975(1969).  


\bibitem{Gribov}
V.N. Gribov,  {\it ``Inelastic processes at super high-energies and the problem of nuclear cross-sections,''}
  Sov.\ J.\ Nucl.\ Phys.\  {\bf 9}, 369 (1969)
  [Yad.\ Fiz.\  {\bf 9}, 640 (1969)];\,\,\, {\it ``Space-time description of hadron interactions at high-energies,''}
Proc. ITEP School on Elementary particle physics, v.1, p.65 (1973);   hep-ph/0006158.

\bibitem{GIP} 
  V.~N.~Gribov, B.~L.~Ioffe and I.~Y.~Pomeranchuk,
  {\it ``What is the range of interactions at high-energies,''}
  Sov.\ J.\ Nucl.\ Phys.\  {\bf 2}, 549 (1966)
  [Yad.\ Fiz.\  {\bf 2}, 768 (1965)].
 
\bibitem{Ioffe} 
  B.~L.~Ioffe,
  {\it ``Space-time picture of photon and neutrino scattering and electroproduction cross-section asymptotics,''}
  Phys.\ Lett.\  {\bf 30B}, 123 (1969).
  doi:10.1016/0370-2693(69)90415-8
   \bibitem{KOLEB}
Yuri V. Kovchegov and Eugene Levin, {\it `` Quantum Chromodynamics at High Energies"}, Cambridge Monographs on Particle Physics, Nuclear Physics and Cosmology, Cambridge University Press, 2012 .  
 
    \bibitem{MV}
L. McLerran and R. Venugopalan, 
{\it ``Computing quark and gluon distribution functions for very large nuclei"},
Phys. Rev. {\bf D49} (1994) 2233,
{\it ``Gluon distribution functions for very large nuclei at small transverse momentum"}, Phys. Rev. {\bf D49} (1994), 3352;     {\it `Green?s function in the color field of a large nucleus"}, {\bf D50} (1994) 2225; {\it ``
Fock space distributions, structure functions, higher twists, and small $x$"} ,
 {\bf D59} (1999) 09400.

  \bibitem{MUCD}
 A.~H.~Mueller,
  {\it ``Soft Gluons In The Infinite Momentum Wave Function And The BFKL Pomeron,''}
  Nucl.\ Phys.\  B {\bf 415}, 373 (1994);
 \,\,{\it ``Unitarity and the BFKL pomeron,''}
  Nucl.\ Phys.\  B {\bf 437} (1995) 107
  [arXiv:hep-ph/9408245].

     \bibitem{B}
I.~Balitsky,
{\it ``Operator expansion for high-energy scattering"},
[arXiv:hep-ph/9509348];\,\,
{\it ``Factorization and high-energy effective action"}, {\it Phys.\ Rev.} {\bf D60}, 014020 (1999)
[arXiv:hep-ph/9812311].\,\,\,
\bibitem{K}
Y.~V.~Kovchegov,
{\it ``
Small-x $F_2$
structure function of a nucleus including multiple Pomeron exchanges"'}
{\it Phys.\ Rev.}  {\bf D60}, 034008  (1999),
[arXiv:hep-ph/9901281].
   \bibitem{JIMWLK}
J.~Jalilian-Marian, A.~Kovner, A.~Leonidov, and H.~Weigert, {\it ``The BFKL
  equation from the Wilson renormalization group"}
  \href{http://dx.doi.org/10.1016/S0550-3213(97)00440-9}, Nucl. Phys. {\bf
  B504} (1997)  415--431,
\href{http://arxiv.org/abs/hep-ph/9701284}[ arXiv:hep-ph/9701284];\,\,
J.~Jalilian-Marian, A.~Kovner, A.~Leonidov, and H.~Weigert, {\it ``The Wilson
  renormalization group for low x physics: Towards the high density regime"}
  \href{http://dx.doi.org/10.1103/PhysRevD.59.014014}, Phys.Rev. {\bf D59}
  (1998)  014014,
\href{http://arxiv.org/abs/hep-ph/9706377}[arXiv:hep-ph/9706377
  [hep-ph]];\,\,\,
A.~Kovner, J.~G. Milhano, and H.~Weigert, {\it ``Relating different approaches to
  nonlinear QCD evolution at finite gluon density"}
  \href{http://dx.doi.org/10.1103/PhysRevD.62.114005}, Phys. Rev. {\bf
  D62} (2000)  114005,
\href{http://arxiv.org/abs/hep-ph/0004014}[ arXiv:hep-ph/0004014];\,\,\,
E.~Iancu, A.~Leonidov, and L.~D. McLerran, {\it Nonlinear gluon evolution in the
  color glass condensate. I"}
  \href{http://dx.doi.org/10.1016/S0375-9474(01)00642-X},Nucl. Phys. {\bf
  A692} (2001)  583--645,
\href{http://arxiv.org/abs/hep-ph/0011241}[ arXiv:hep-ph/0011241];\,\,\,
E.~Iancu, A.~Leonidov, and L.~D. McLerran, {\it``The renormalization group
  equation for the color glass condensate"}
  \href{http://dx.doi.org/10.1016/S0370-2693(01)00524-X}, Phys. Lett. {\bf
  B510} (2001)  133--144,
\href{http://arxiv.org/abs/hep-ph/0102009}[ arXiv:hep-ph/0102009];\,\,
E.~Ferreiro, E.~Iancu, A.~Leonidov, and L.~McLerran, {\it ``Nonlinear gluon
  evolution in the color glass condensate. II"}
  \href{http://dx.doi.org/10.1016/S0375-9474(01)01329-X}, Nucl. Phys. {\bf
  A703} (2002)  489--538,
\href{http://arxiv.org/abs/hep-ph/0109115}[ arXiv:hep-ph/0109115];\,\,\,
H.~Weigert,
{\it Unitarity at small Bjorken $ x$},  Nucl.\ Phys.  {\bf A703}, 823 (2002),
[arXiv:hep-ph/0004044].
 
\bibitem{GIJMV}
F.~Gelis, E.~Iancu, J.~Jalilian-Marian and R.~Venugopalan,
{\it ``The Color Glass Condensate,''}
Ann. Rev. Nucl. Part. Sci. \textbf{60} (2010), 463-489
doi:10.1146/annurev.nucl.010909.083629
[arXiv:1002.0333 [hep-ph]].

   \bibitem{BFKL}
V.~S. Fadin, E.~A. Kuraev and L.~N. Lipatov,
{\it ``On the pomeranchuk singularity in asymptotically free theories"},
\newblock Phys. Lett. {\bf B60}, 50 (1975);\,\,\,
E.~A. Kuraev, L.~N. Lipatov and V.~S. Fadin,
{\it``The Pomeranchuk Singularity in Nonabelian Gauge Theories"}
\newblock Sov. Phys. JETP {\bf 45}, 199 (1977),
\newblock [Zh. Eksp. Teor. Fiz.72,377(1977)];\,\,\,
{\it ``The Pomeranchuk Singularity in Quantum Chromodynamics,''}
I.~I. Balitsky and L.~N. Lipatov,
\newblock Sov. J. Nucl. Phys. {\bf 28}, 822 (1978),
\newblock [Yad. Fiz.28,1597(1978)].
\bibitem{LI}
  L.~N.~Lipatov,
  {\it ``The Bare Pomeron in Quantum Chromodynamics,''}
  Sov.\ Phys.\ JETP {\bf 63}, 904 (1986)
  [Zh.\ Eksp.\ Teor.\ Fiz.\  {\bf 90}, 1536 (1986)].

\bibitem{GLR} 
  L.~V.~Gribov, E.~M.~Levin and M.~G.~Ryskin,
  {\it ``Semihard Processes in QCD,''}
  Phys.\ Rept.\  {\bf 100}, 1 (1983).
  doi:10.1016/0370-1573(83)90022-4
  \bibitem{GLR1}
E.~M.~Levin and M.~G.~Ryskin,
  {\it ``High-energy hadron collisions in QCD,''}
  Phys.\ Rept.\  {\bf 189}, 267 (1990).
  \bibitem{MUQI}
A. H. Mueller and J. Qiu, 
{\it ``  Gluon recombination and shadowing at small values of $x$",} Nucl. Phys. {\bf B268} (1986) 427.
  \bibitem{MUPA}
A. H. Mueller and B. Patel,
{\it ``Single and double BFKL pomeron exchange and a dipole picture of high-energy hard processes",}
Nucl. Phys. {\bf B425} (1994) 471.  
\bibitem{BART}
J.~Bartels, M.~Braun and G.~Vacca,
{\it ``Pomeron vertices in perturbative QCD in diffractive scattering,''}
Eur. Phys. J. C \textbf{40} (2005), 419-433
doi:10.1140/epjc/s2005-02152-x
[arXiv:hep-ph/0412218 [hep-ph]];\,\,
J.~Bartels and C.~Ewerz,
{\it ``Unitarity corrections in high-energy QCD,''}
JHEP \textbf{09} (1999), 026
doi:10.1088/1126-6708/1999/09/026
[arXiv:hep-ph/9908454 [hep-ph]];\,\,\,
 J.~Bartels and M.~Wusthoff,
{\it ``The Triple Regge limit of diffractive dissociation in deep inelastic scattering,''}
Z. Phys. C \textbf{66} (1995), 157-180
doi:10.1007/BF01496591;
J.~Bartels,
{\it ``Unitarity corrections to the Lipatov pomeron and the four gluon operator in deep inelastic scattering in QCD,''}
Z. Phys. C \textbf{60} (1993), 471-488
doi:10.1007/BF01560045

 \bibitem{BRN}
M.~Braun,
{\it ``Conformal invariant pomeron interaction in the perurbative QCD with large  $N_c$,''}
Phys. Lett. B \textbf{632} (2006), 297-304
doi:10.1016/j.physletb.2005.10.054
[arXiv:hep-ph/0512057 [hep-ph]]; \,\,
{\it ``Nucleus nucleus interaction in the perturbative QCD,''}
Eur. Phys. J. C \textbf{33} (2004), 113-122
doi:10.1140/epjc/s2003-01565-9
[arXiv:hep-ph/0309293 [hep-ph]];
{\it ``Nucleus-nucleus scattering in perturbative QCD with $N_c  \to$ infinity,''}
Phys. Lett. B \textbf{483} (2000), 115-123
doi:10.1016/S0370-2693(00)00571-2
[arXiv:hep-ph/0003004 [hep-ph]];\,\,
{\it ``Structure function of the nucleus in the perturbative QCD with $N_c \to$ infinity (BFKL pomeron fan diagrams),''}
Eur. Phys. J. C \textbf{16} (2000), 337-347
doi:10.1007/s100520050026
[arXiv:hep-ph/0001268 [hep-ph]];\,\,
{\it``The system of four reggeized gluons and the three-pomeron vertex in the high colour limit"}
Eur.  Phys. J. {\bf C6}, 321 (1999) [arXiv:hep-ph/9706373];\,\,\
M.~Braun and G.~Vacca,
{\it ``Triple pomeron vertex in the limit  $N_c \to$ infinity,''}
Eur. Phys. J. C \textbf{6} (1999), 147-157
doi:10.1007/s100520050328
[arXiv:hep-ph/9711486 [hep-ph]].



\bibitem{KOLE}
Y.~V.~Kovchegov and E.~Levin,
{\it ``Diffractive dissociation including multiple pomeron exchanges in high parton density QCD,''}
Nucl. Phys. B \textbf{577} (2000), 221-239
doi:10.1016/S0550-3213(00)00125-5
[arXiv:hep-ph/9911523 [hep-ph]].

\bibitem{LELU1}
E. Levin and M. Lublinsky,
  {\it ``Towards a symmetric approach to high energy evolution: Generating
  functional with Pomeron loops,''}
  Nucl.\ Phys.\  A {\bf 763} (2005) 172
  [arXiv:hep-ph/0501173].
  \bibitem{LELU2}
E. Levin and M. Lublinsky,  
 {\it ``Balitsky's hierarchy from Mueller's dipole model and more about target correlations,''}
  Phys.\ Lett.\  B {\bf 607} (2005) 131
  [arXiv:hep-ph/0411121];\,\, 
  {\it ``A linear evolution for non-linear dynamics and correlations in  realistic nuclei,''}
  Nucl.\ Phys.\  A {\bf 730} (2004) 191
  [arXiv:hep-ph/0308279].
\bibitem{LMP}
E. Levin, J. Miller and A. Prygarin,
  {\it ``Summing Pomeron loops in the dipole approach,''}
  Nucl.\ Phys.\  {\bf A806 } (2008)  245,
  [arXiv:0706.2944 [hep-ph]].
\bibitem{AKLL}
  T.~Altinoluk, C.~Contreras, A.~Kovner, E.~Levin, M.~Lublinsky and A.~Shulkim,
  {\it ``QCD reggeon calculus from JIMWLK Evolution,''}
  Int.\ J.\ Mod.\ Phys.\ Conf.\ Ser.\  {\bf 25} (2014) 1460025;\,\,\,
   T.~Altinoluk, N.~Armesto, A.~Kovner, E.~Levin and M.~Lublinsky,
  {\it ``KLWMIJ Reggeon field theory beyond the large $ N_{c}$ limit,''}
  JHEP {\bf 1408} (2014) 007.
  \bibitem{AKLL1}
 T.~Altinoluk, A.~Kovner, E.~Levin and M.~Lublinsky,
  {\it ``Reggeon Field Theory for Large Pomeron Loops,''}
  JHEP {\bf 1404} (2014) 075
  [arXiv:1401.7431 [hep-ph]].;\,\,\, 
  T.~Altinoluk, C.~Contreras, A.~Kovner, E.~Levin, M.~Lublinsky and A.~Shulkin,
  {\it ``QCD Reggeon Calculus From KLWMIJ/JIMWLK Evolution: Vertices, Reggeization and All,''}
  JHEP {\bf 1309} (2013) 115.
\bibitem{LEPP}
E.~Levin,
  {\it ``Dipole-dipole scattering in CGC/saturation approach at high energy: summing Pomeron loops,'}
  JHEP {\bf 1311} (2013) 039
  [arXiv:1308.5052 [hep-ph]].
  
\bibitem{KLW}
A.~Kovner, M.~Lublinsky and U.~Wiedemann,
{\it ``From bubbles to foam: Dilute to dense evolution of hadronic wave function at high energy,''}
JHEP \textbf{06} (2007), 075
doi:10.1088/1126-6708/2007/06/075
[arXiv:0705.1713 [hep-ph]].
\bibitem{AKLP}
T.~Altinoluk, A.~Kovner, M.~Lublinsky and J.~Peressutti,
{\it ``QCD Reggeon Field Theory for every day: Pomeron loops included,''}
JHEP \textbf{03} (2009), 109
doi:10.1088/1126-6708/2009/03/109
[arXiv:0901.2559 [hep-ph]].

  \bibitem{RY}
I. Gradstein and I. Ryzhik, {\it  Table of Integrals, Series, and Products},
Fifth Edition, Academic Press, London, 1994.

      \bibitem{MPSI}
A.~H.~Mueller and G.~Salam,
{\it ``Large multiplicity fluctuations and saturation effects in onium collisions,''}
Nucl. Phys. B \textbf{475} (1996), 293-320
doi:10.1016/0550-3213(96)00336-7
[arXiv:hep-ph/9605302 [hep-ph]];\,\,\,
G.~Salam,
{\it ``Studies of unitarity at small x using the dipole formulation,''}
Nucl. Phys. B \textbf{461} (1996), 512-538
doi:10.1016/0550-3213(95)00658-3
[arXiv:hep-ph/9509353 [hep-ph]];\,\,\,
E.~Iancu and A.~Mueller,
{\it ``Rare fluctuations and the high-energy limit of the S matrix in QCD,''}
Nucl. Phys. A \textbf{730} (2004), 494-513
doi:10.1016/j.nuclphysa.2003.10.019
[arXiv:hep-ph/0309276 [hep-ph]];
\,\,{\it ``From color glass to color dipoles in high-energy onium onium scattering,''}
Nucl. Phys. A \textbf{730} (2004), 460-493
doi:10.1016/j.nuclphysa.2003.10.017
[arXiv:hep-ph/0308315 [hep-ph]].
\bibitem{KLL}
   A.~Kovner, E.~Levin and M.~Lublinsky,
{\it ``QCD unitarity constraints on Reggeon Field Theory,''}
JHEP \textbf{08} (2016), 031
doi:10.1007/JHEP08(2016)031
[arXiv:1605.03251 [hep-ph]].

    \bibitem{MUDIA}
A. H. Mueller,
  {\it ``O(2,1) analysis of single particle spectra at high energy,''}
 Phys. Rev. {\bf D2} (1970) 2963.
   \bibitem{KTINC}
   Y.~V.~Kovchegov and K.~Tuchin,
  {\it ``Inclusive gluon production in DIS at high parton density,''}
  Phys.\ Rev.\   {\bf D65} (2002) 074026
  [arXiv:hep-ph/0111362].
  \bibitem{CMSMULT}
V.~Khachatryan \textit{et al.} [CMS],
{\it ``Charged Particle Multiplicities in $pp$ Interactions at $\sqrt{s}=0.9$, 2.36, and 7 TeV,''}
JHEP \textbf{01} (2011), 079
doi:10.1007/JHEP01(2011)079
[arXiv:1011.5531 [hep-ex]].
  
\bibitem{GKLM}
E.~Gotsman, A.~Kormilitzin, E.~Levin and U.~Maor,
{\it ``QCD motivated approach to soft interactions at high energies: nucleus-nucleus and hadron-nucleus collisions,''}
Nucl. Phys. A \textbf{842} (2010), 82-101
doi:10.1016/j.nuclphysa.2010.04.016
[arXiv:0912.4689 [hep-ph]].
  
  \bibitem{LLN}
A.~Likhoded, A.~Luchinsky and A.~Novoselov,
{\it``Light hadron production in inclusive pp-scattering at LHC,''}
Phys. Rev. D \textbf{82} (2010), 114006
doi:10.1103/PhysRevD.82.114006
[arXiv:1005.1827 [hep-ph]].

\bibitem{KAPO}
A.~Kaidalov and M.~Poghosyan,
{\it ``Predictions of Quark-Gluon String Model for pp at LHC,''}
Eur. Phys. J. C \textbf{67} (2010), 397-404
doi:10.1140/epjc/s10052-010-1301-y
[arXiv:0910.2050 [hep-ph]].

\bibitem{LERE}
E.~Levin and A.~H.~Rezaeian,
{\it ``Gluon saturation and inclusive hadron production at LHC,''}
Phys. Rev. D \textbf{82} (2010), 014022
doi:10.1103/PhysRevD.82.014022
[arXiv:1005.0631 [hep-ph]].
\bibitem{KLN}D.~Kharzeev and M.~Nardi,\textit{
  ``Hadron production in nuclear collisions at RHIC and high density QCD,''}
 Phys.~Lett.~B
\textbf{507}, 121 (2001) {[}nucl-th/0012025{]}.
\bibitem{KLN} . 
 D.~Kharzeev and E.~Levin, \textit{` `Manifestations of high density
QCD in the first RHIC data,''} Phys.\ Lett.\ B \textbf{523} (2001)
79, {[}nucl-th/0108006{]};\,\, D.~Kharzeev, E.~Levin and M.~Nardi,
\textit{``The Onset of classical QCD dynamics in relativistic heavy
ion collisions,''} Phys.\ Rev.\ C \textbf{71} (2005) 054903, {[}hep-ph/0111315{]};
\textit{``Hadron multiplicities at the LHC,''} J.\ Phys.\ G \textbf{35}
(2008) no.5, 054001.38 {[}arXiv:0707.0811 {[}hep-ph{]}{]}. 


\bibitem{DKLN} A.~Dumitru, D.~E.~Kharzeev, E.~M.~Levin and Y.~Nara,
\textit{`` ``Gluon Saturation in $pA$ Collisions at the LHC: KLN
Model Predictions For Hadron Multiplicities,''} Phys.\ Rev.\ C
\textbf{85} (2012) 044920 {[}arXiv:1111.3031 {[}hep-ph{]}{]}





\bibitem{LAPPI}T.~Lappi, \textit{ ``Energy dependence of the saturation scale and the charged multiplicity in pp and AA collisions,''}
Eur.~Phys.~J.~C \textbf{71},
1699 (2011) {[}arXiv:1104.3725 {[}hep-ph{]}{]}.

\bibitem{MULIB}
A.~H.~Mueller,
{\it ``Toward equilibration in the early stages after a high-energy heavy ion collision,''}
Nucl. Phys. B \textbf{572} (2000), 227-240
doi:10.1016/S0550-3213(99)00502-7
[arXiv:hep-ph/9906322 [hep-ph]].

    
  \bibitem{LHPD} 
  Y.~L.~Dokshitzer, V.~A.~Khoze, S.~I.~Troian and A.~H.~Mueller,
  {\it ``QCD Coherence in High-Energy Reactions,''}
  Rev.\ Mod.\ Phys.\  {\bf 60}, 373 (1988).
  doi:10.1103/RevModPhys.60.373
   \bibitem{KNO} 
Z.~Koba, H.~B.~Nielsen and P.~Olesen,
{\it``Scaling of multiplicity distributions in high-energy hadron collisions,''}
Nucl. Phys. B \textbf{40} (1972), 317-334
doi:10.1016/0550-3213(72)90551-2

\bibitem{GRIBPC}
V. ~N. Gribov,
  {\it ``A reggeon diagram technique,''}
   Sov.\ Phys.\ JETP {\bf 26} (1967) 414
  [ Zh.\ Eksp.\ Teor.\ Fiz.\  {\bf 53} (1967) 654].
 \bibitem{ABKA}
  V.~Abramovskii and O.~Kancheli,
{\it ``Regge branching and distribution of hadron multiplicity at high energies,''}
Pisma Zh. Eksp. Teor. Fiz. \textbf{15} (1972), 559-563.
  \bibitem{MAWA}
S.~G.~Matinyan and W.~Walker,
{\it ``Multiplicity distribution and mechanisms of the high-energy hadron collisions,''}
Phys. Rev. D \textbf{59} (1999), 034022
doi:10.1103/PhysRevD.59.034022
[arXiv:hep-ph/9801219 [hep-ph]] and reference therein.
 \bibitem{PDG}
C. Patrignani et al. (Particle Data Group), Chin. Phys. C, 40, 100001 (2016).  
 
 \end{thebibliography}
\end{document}